\documentclass[preprint,12pt]{aastex}

\newcommand{\unit}[1]{\; \mbox{#1}}
\newcommand{\un}[1]{_{\mbox{\scriptsize #1}}}
\usepackage{epsfig}
\usepackage{amssymb, color}

\begin{document}
\title{Antiprotons from spallations of cosmic rays on
interstellar matter}
\author{F. Donato\altaffilmark{1}}
\affil{Laboratoire de Physique Th\'eorique {\sc lapth},
Annecy--le--Vieux, 74941, France}
\email{donato@lapp.in2p3.fr}
\author{D. Maurin and P. Salati\altaffilmark{2}}
\affil{Laboratoire de Physique Th\'eorique {\sc lapth},
Annecy--le--Vieux, 74941, France \\
Universit\'e de Savoie, Chamb\'ery, 73011, France}
\email{maurin@lapp.in2p3.fr, salati@lapp.in2p3.fr}
\author{A. Barrau and G. Boudoul}
\affil{Institut des Sciences Nucl\'eaires {\sc isn},
Grenoble, 38026, France \\
Universit\'e Joseph Fourier, St Martin d'H\`eres, 38400, France}
\email{barrau@isn.in2p3.fr, boudoul@isn.in2p3.fr}
\and
\author{R. Taillet}
\affil{Laboratoire de Physique Th\'eorique {\sc lapth},
Annecy--le--Vieux, 74941, France \\
Universit\'e de Savoie, Chamb\'ery, 73011, France.}
\email{taillet@lapp.in2p3.fr}
\altaffiltext{1}{{\sc infn} post--doctoral Fellow}
\begin{abstract}

Cosmic ray antiprotons provide an important probe for the study of
the galactic Dark Matter, as they could be produced by neutralino
annihilations, primordial black holes evaporations or other
exotic sources. On the other hand,
antiprotons are anyway produced by standard nuclear reactions
of cosmic ray nuclei on interstellar matter ({\em spallations}),
that are known to occur in the Galaxy.
This process is responsible for a background flux
that must be carefully determined to estimate the detectability of
an hypothetical exotic signal.

In this paper we provide a new
evaluation of the interstellar cosmic antiproton flux that is fully consistent
with cosmic ray nuclei in the framework of a two-zone diffusion model.  
We also study and conservatively quantify all possible sources 
of uncertainty that may affect that antiproton flux. In particular, 
the primary cosmic rays are by now so well measured that the 
corresponding error is removed. Uncertainties related to propagation 
are shown to range between 10\%
and 25\%, depending on which part of the spectrum is considered.

\end{abstract}

%%%%%%%%%%%%%%%%%%%%%%%%%%%%%%%%%%%%%%%%%%%%%%%%%%%%%%%%%%%%%%%%%%%%%%%%%%%%

\section{Introduction}
The study of the cosmic ray antiproton spectrum has been a great challenge
since the first measurements made at the end of the seventies.
Actually, the first experiments provided data which, in the low energy tail, 
showed some excess  when compared to the current model predictions.
This discrepancy stimulated a great interest into alternative explanations, viz
the possible existence of primary antiproton sources. Such an interest did not
fade even when further experimental data seemed to agree with theoretical 
predictions in standard leaky box models (see for example Stephens 
\& Golden 1988 and references therein). 

Various primary antiproton sources have been proposed
(Silk \& Srednicki 1984; Stecker, Rudaz, \& Walsh 1985;
Ellis et al. 1988; Starkman \& Vachaspati 1996; Mitsui, Maki, \&
Orito 1996).
The case of supersymmetric sources -- relic neutralinos in the galactic 
halo --  has received a particular attention
and constraints on {\sc susy} parameters have been investigated
by comparing experimental data to theoretical predictions
(Bottino et al. 1995; Chardonnet et al. 1996; Bottino et al. 1998; Bergstr\"om,
Edsj\"{o}, \& Ullio 1999). However, an important problem with this comparison 
is that an accurate estimation of the background secondary antiproton flux
produced by spallations is mandatory. 

In this paper, we focus on this secondary antiproton
flux, which we will call ``background" antiproton flux, having in mind
the possibility of using it to determine whether one of the  primary
components (``signal") discussed above could be seen against it or not
(P. Salati \& al, in preparation; A. Barrau \& al, in preparation).
Such  hypothetical signals will not be further discussed in this paper.
We believe that now is a good time for a detailed evaluation of the
background flux, since the next measurements of $\bar{p}$ spectra
should be very accurate at low energy ($\sim 100\unit{MeV} - 10\unit{GeV}$)
especially in the forthcoming ten years ({\sc ams}, {\sc bess},
{\sc pamela}, \dots).
On the theoretical side, progress has already been made in many directions.
Here are some milestones on the way: 
(i) the inelastic non-annihilating cross-section for $\bar{p}$
(Tan \& Ng 1982, 1983), giving rise to the so--called tertiary
contribution, has been taken into account
(ii) the $p + He\un{\sc ism}\rightarrow \bar{p}$ contribution
has been considered by means of a simple geometric approach
(Gaisser \& Schaefer 1992),
(iii) reacceleration has been considered (Simon \& Heinbach 1996),
(iv) propagation has been modeled in a more realistic two--zone
diffusion model (Halm, Jansen, \& de Niem 1993; Chardonnet et al. 1996),
(v) the $(p,He)+(H,He)\un{\sc ism}$ reactions have been re-estimated
in a more sophisticated nuclear
Monte Carlo (Simon, Molnar, \& Roesler 1998),
(vi) the great variety of cosmic rays has been treated in a more coherent way
(Moskalenko, Strong \& Reimer 1998).
As far as we know, all these ingredients have only been considered
simultaneously in Moskalenko et al. 1998
(see also Moskalenko et al. 2001).

We propose to go beyond this type of study and to use the results of 
our systematic analysis of nuclei (Maurin et al. 2001,
referred hereafter as Paper I) to ascertain the theoretical uncertainties
on the interstellar secondary antiproton energy spectrum.
This goal has never been achieved before, even in Moskalenko et al. 1998, 2001.
The paper is organized as follows.
Separate sections are devoted to all the ingredients entering the
calculation of the $\bar{p}$ background: measured H
and He flux, secondary production, tertiary contribution and
propagation. Whithin each section, we first discuss the model used
and the associated parameters;
then we estimate the uncertainty they induce in the $\bar{p}$ background.
An important aspect is worth a warning at this point. As will be
discussed in section~\ref{sec:results}, 
the effect of solar modulation may be decoupled from the problem of 
interstellar propagation and this problem will not be addressed here.
When a modulated flux is needed, we will use a simple force-field approximation
modulation scheme, as in most cosmic antiprotons studies.
Would a more careful treatment of solar modulation be needed (see for
example Bieber et al. 1999), an interstellar flux can easily be 
obtained by demodulation (the force-field approximation
modulation scheme is reversible). This interstellar flux could then be
used as an input for any other preferred treatment of solar 
modulation.

To sum up, we used results from a systematic nuclei cosmic
ray analysis to consistently derive an antiproton secondary flux in the 
framework of diffusion models. 
As an important consequence we could study and quantify most of the 
uncertainties: in the propagation, in the nuclear physics and in 
the primary cosmic ray. 
We feel that our results will be valuable not only for speculations on
primary contributions to that flux but also for the experimental groups
which are going to perform very accurate antiproton measurements in
the near future.

%%%%%%%%%%%%%%%%%%%%%%%%%%%%%%%%%%%%%%%%%%%%%%%%%%%%%%%%%%%%%%%%%%%%%%%%%%%%
\section{Proton and Helium primary spectra}
\label{section:trop bien}
The secondary antiprotons are yielded by the spallation of cosmic ray nuclei
over the interstellar medium (see Appendix A for the formul\ae).
The most abundant species in cosmic rays are protons and helium,
and the contribution of heavier nuclei to the antiproton production
is negligible.
Until recently, their spectra were known with a modest accuracy and
the data from different experiments were often
incompatible at high energy. This induced an uncertainty of
some tens of percents in the predicted antiproton spectrum.
Recent measurements made by the balloon--borne spectrometer {\sc bess}
(Sanuki et al. 2000) and by the {\sc ams} detector during the space
shuttle flight
(Alcaraz et al. 2000a, 2000b, 2000c) dramatically reduced the
uncertainties both on  proton and helium spectra.
We fitted the high energy (T $>$ 20 GeV/n) part of these measured spectra
with the power law:
\begin{equation}
\Phi (\rm T) = N\,(\rm T/\unit{GeV/n})^{-\gamma}\;\;,
\label{eq:cosmic}
\end{equation}
where the kinetic energy per nucleon T is given in units of GeV/n and the
normalization factor N in units of $\unit{m}^{-2} \unit{s}^{-1}
\unit{sr}^{-1}\unit{(GeV/n)}^{-1}$.
This provides a good description down to
the threshold energy for the  antiproton production.

\begin{figure}[p]
\centerline{\includegraphics*[bb=55 118 530 660,clip,width=1.
         \textwidth]{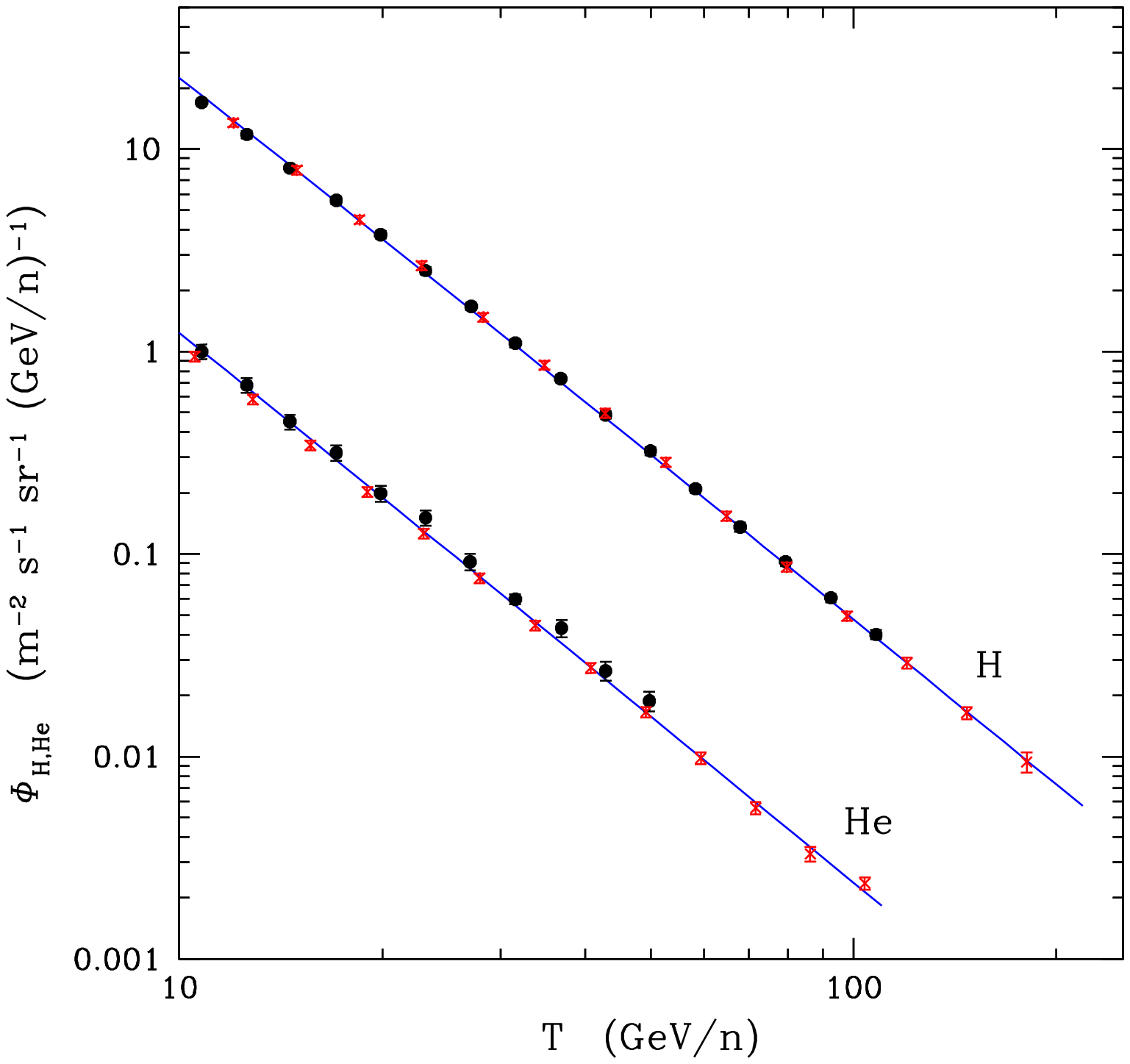}}
\caption{The upper (lower) curve displays the measured proton (helium)
flux along with an analytical fit (see text). On both curves, data
are from {\sc ams} (Alcaraz et al. 2000a, 2000b, 2000c)
(crosses) and {\sc bess} (Sanuki et al. 2000) (filled circles).}
\label{fig:proton_helium}
\end{figure}

We fitted the {\sc bess} and {\sc ams} data both separately and combined,
obtaining very similar results.
This is obvious since the data from the two experiments are now totally
compatible, as can be  seen in Fig. \ref{fig:proton_helium}. 
The upper curve presents our fit on the
combined proton data. The best fit corresponds to
$N = 13249\unit{m}^{-2} \unit{s}^{-1}
\unit{sr}^{-1}\unit{(GeV/n)}^{-1}$ and
$\gamma = 2.72$. We do not plot the spectra obtained from the best fits
on the single {\sc bess} and {\sc ams} data because of their complete
overlap with the plotted curve. We did the same for helium (lower curve)
and the corresponding numbers are $N = 721\unit{m}^{-2} \unit{s}^{-1}
\unit{sr}^{-1}\unit{(GeV/n)}^{-1}$ and $\gamma = 2.74$.
The 1--$\sigma$ deviation from the best fit spectrum does not exceed 
1$\%$ for both species.
Consequently, the corresponding  uncertainty on the antiproton spectrum 
is smaller than the ones discussed in the next sections, and it will be
neglected in the rest of this paper.
The situation has significantly improved since Bottino \& al. 1998, where
an error of $\pm$ 25\% was quoted.

%%%%%%%%%%%%%%%%%%%%%%%%%%%%%%%%%%%%%%%%%%%%%%%%%%%%%%%%%%%%%%%%%%%%%%%%%%%%
\section{Antiprotons production: secondary sources}
\label{section:ici}
Whereas p--p interactions are clearly the dominant process for secondary
antiproton production in the galaxy, it has been realized long ago that
  p--nucleus and nucleus--nucleus collisions should also be taken into account
\cite{gaisser}. They not only enhance the antiproton flux as a whole but also
change its low energy tail, mostly for kinematical reasons. 
Unfortunately, very few experimental data are available on antiproton production 
cross-sections  in nuclear collisions. A model-based evaluation is therefore 
necessary,
and we chose to use  the {\sc dtunuc} program. We first discuss sub-threshold 
antiproton production. Then we present the results of our calculations of 
above-threshold production, which we compare to experimental data and 
analytical formul\ae.

                         %#####################%
\subsection{p--p interaction}
\label{pp_interaction}
Antiproton production {\it via} the proton--proton interaction is the first
reaction that one has to take into account in order to evaluate the 
$\bar{p}$ flux.
So far, the Tan \& Ng parameterization of $\bar{p}$ cross-section
(Tan \& Ng 1982, 1983) has been used by almost all studies on cosmic ray
antiprotons.
To be more precise, we recall the form of secondary contribution ({\it e.g.}
eq. [\ref{SECONDARY_1}], Appendix \ref{sec:HEL})
\begin{equation}
q_{\bar{p}}^{sec}(r,E) \; = \;
{\displaystyle \int\un{Threshold}^{\infty}} \,
{\displaystyle \frac{d \sigma}{dE}}
\left\{ p(E') + H\un{\sc ism} \rightarrow \bar{p}(E) \right\} \, n_{He} \,
\left\{ 4 \, \pi \, \Phi_{p}(r,E') \right\} \, dE'\;\;.
\end{equation}
Thus, in order to evaluate the secondary contribution of $p-H\un{\sc ism}$
reaction, we used the parameterization of Tan \& Ng (1982, 1983). 
We refer the interested reader to the short discussion 
in Bottino et al. (1998) for further details, or to the source
papers (Tan \& Ng 1982, 1983) for a complete description.
Finally, as 
an illustration,
the impact of kinematics and threshold for the production rate can be found
in Gaisser \& Schaefer (1992).
                         %#####################%
\subsection{Calculation of the differential cross-section of antiprotons
production in p-He, He-p and He-He reactions}
\label{ab_interaction}
Some discrepancies between simple scalings
of p--p cross-sections and experimental data on p--nucleus antiproton 
production
cross-sections near threshold have been explained by taking into 
account internal
nuclear Fermi motion \cite{shor}. We first show that this effect does not
change the cosmic antiproton spectrum.
In such models, the momentum distribution is
described by a double-gaussian function normalized to the total 
number of nucleons.
The parameters are determined from scattering experiments \cite{moniz}
and simple scaling laws. The cross-section results from a convolution
\begin{equation}
\frac{d^2\sigma_{p+nucleus\rightarrow\bar{p}+X}}{d\Omega dp}=\int d^3 {\bf p_c}
f({\bf p_c}) \frac{d^2\sigma_{N+N\rightarrow\bar{p}+X}}{d\Omega 
dp}(E_{cm})\;\;,
\end{equation}
where ${\bf p_c}$ is the internal nuclear momentum of the target nucleon, N
denotes either a proton or a neutron (the model is isospin 
independent) and E$\un{cm}$ is
the center of mass energy (with an off--shell target nucleon).

Near threshold,
the nucleon--nucleon cross-section can be estimated from
the transition matrix element and the available phase space
by Fermi's golden rule. Using
this simple approach with only one free parameter (namely the matrix 
element), fitted on data, we have been
able to reproduce very well most
experimental results available on subthreshold antiproton production. 
The kinematical term was computed using a Monte
Carlo multi-particle
weighted event according to Lorentz-invariant Fermi phase space, whereas the
integral was performed by adaptable gaussian quadrature.
This method is not relevant to accurately determine the p--He, He--p or
He--He cross-sections at any
energy (as the momentum distribution becomes a $\delta$ function when the
involved momenta are much greater than the Fermi momentum) but just
to investigate their behaviour below the 6 GeV kinetic energy threshold.
The main result is a very fast drop below the threshold. Even after
convolution with the $\approx E^{-2.7}$ differential power law spectrum
of primary cosmic rays, two orders of magnitude are lost in less than
two GeV below the threshold. As a consequence, the subthreshold cross
section can be neglected to compute the secondary antiprotons
flux. The above-threshold discrepancies between data and simple models
cannot be accounted for by
this effect and a numerical Monte Carlo approach is necessary.

Following Simon et al. (1998), the Monte Carlo program
{\sc dtunuc}\footnote{http://sroesler.home.cern.ch/sroesler/}
version 2.3 was therefore used to evaluate the
cross-sections for p--He, He--p and He-He antiproton production reactions.
The p--p reaction can
be well accounted by the Tan \& Ng parameterization
(see previous section)
whereas those involving nuclei
heavier than helium are negligible due to cosmic abundances.
This program is an implementation of the two-component Dual Parton
Model (Capella et al. 1994) based on the Gribov-Glauber approach 
treating soft and hard scattering
processes in
a unified way. Soft processes are parameterized according to Regge
phenomenology
whereas lowest order perturbative {\sc qcd} is used to simulate the 
hard component
\cite{roesler}. This program calls {\sc phojet} \cite{engel}
to treat individual hadron/nucleon/photon-nucleon interaction,
{\sc pythia} \cite{sjostrand} for fragmentation of parton (according to
the Lund model) and {\sc lepto} \cite{lepto} for deep inelastic scattering
off nuclei.
                         %################%
\subsubsection{Comparison with experimental data}
\label{Aurelien deconne}
\begin{figure}[p]
$$
\epsfxsize=17cm
\epsfysize=16cm
\epsfbox{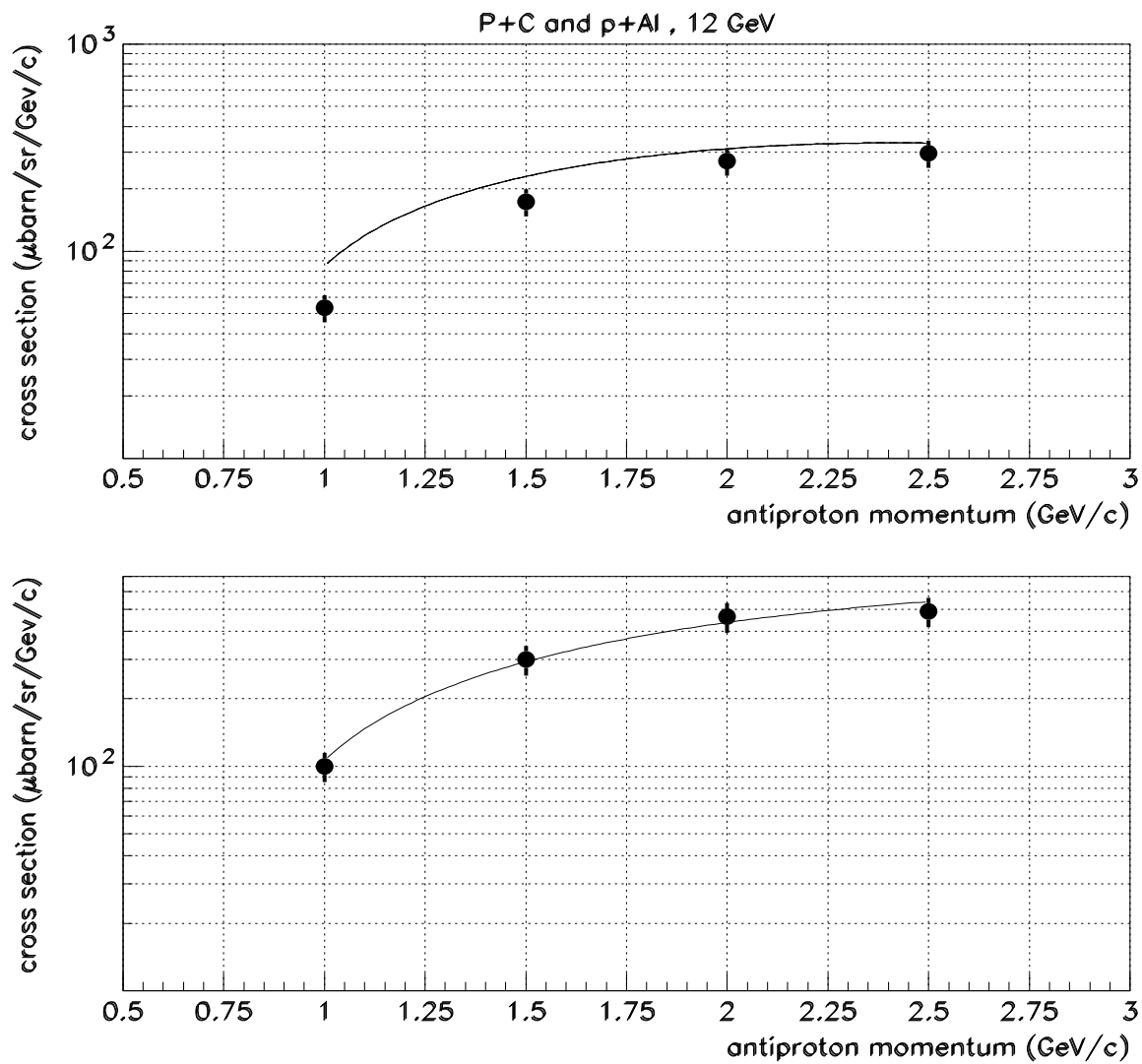}
$$
\caption{Here are displayed the antiproton production cross-section
in p+C (top) and p+Al (bottom) collisions at 12 GeV laboratory kinetic
energy. Filled circles are experimental data \cite{sugaya} and the lines
are from our {\sc dtunuc}
simulations. The error bars have been assumed to be $15\%$.
This value is usual for such experiments and was suggested
by a $\chi^2$ analysis combining most data available.}
\label{fig:sugaya}
\end{figure}

\begin{figure}[p]
$$
\epsfxsize=17cm
\epsfysize=16cm
\epsfbox{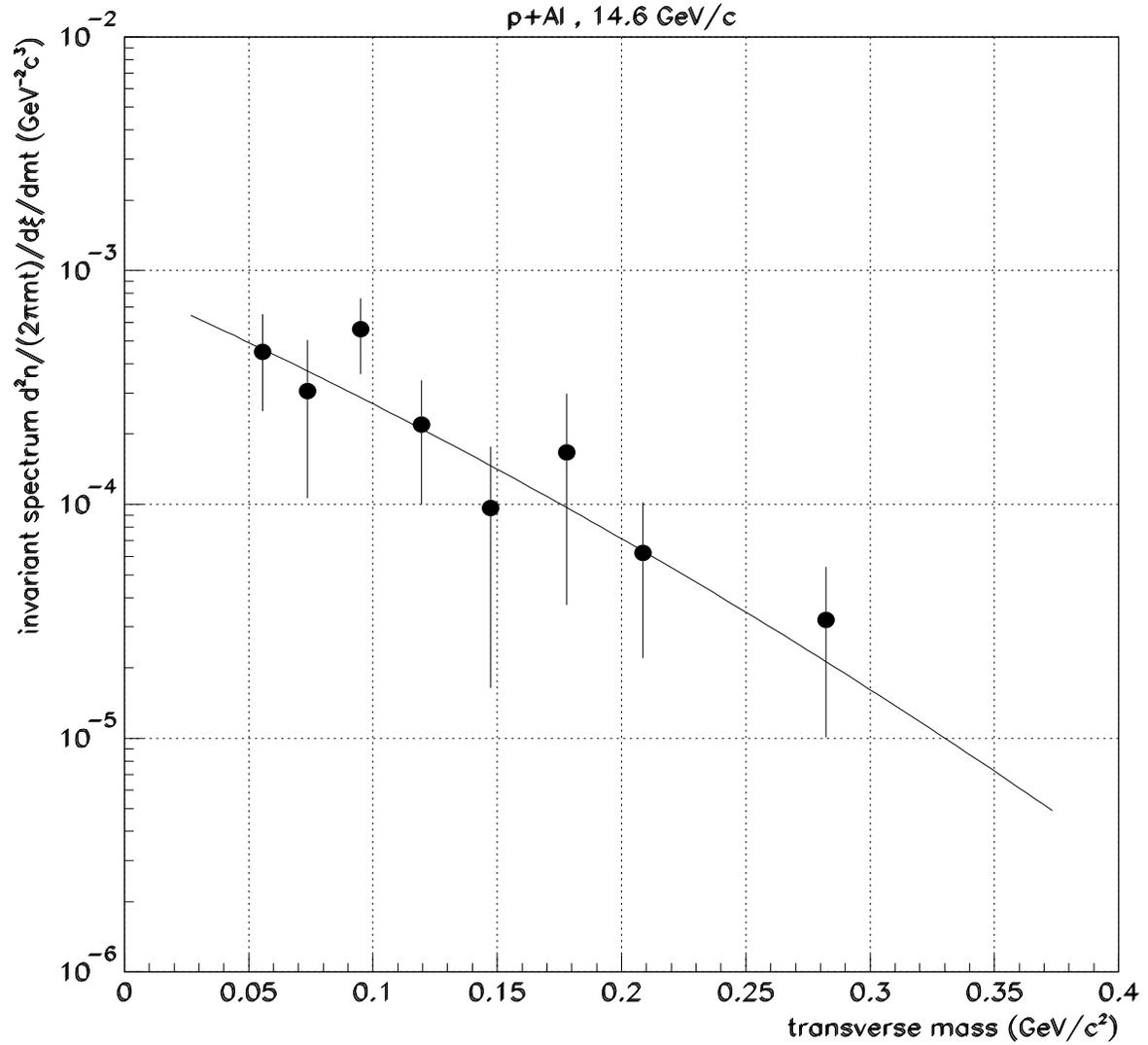}
$$
\caption{Invariant spectrum of $\bar{p}$ in p+Al collisions at 14.6 GeV
laboratory momentum. Filled circles are experimental data (Abbott et
al. 1993) and
the line is from our {\sc dtunuc} simulation.}
\label{fig:abbott}
\end{figure}

The resulting cross-sections have been compared with experimental data on
proton--nucleus collisions.
Figure \ref{fig:sugaya} shows the differential cross-section of antiprotons
production in p+C and p+Al collisions at 12 GeV laboratory kinetic
energy recently measured at the Proton
Synchrotron in the High Energy Accelerator Research Organisation 
({\sc kek--ps})  for different
antiprotons momenta \cite{sugaya}. In most cases,
measurements and {\sc dtunuc} simulations are compatible within uncertainties.
The discrepancies are, anyway, taken into account in section
\ref{sec:uncertainties} as uncertainties on the computed cross-sections.
Figure \ref{fig:abbott} shows the invariant spectrum of antiprotons in p+Al
collisions at 14.6 GeV/c laboratory momentum as a function of
$m_t-m$ where $m_t=\sqrt{p_t^2+m^2}$ as obtained by experiment 802 at
the Brookhaven Tandem Alternating Gradient Synchrotron ({\sc ags})
\cite{abbott1}.
Data points have been normalized by using the inelastic cross-sections
and plotted for a rapidity interval of $1.0<y<1.6$.
The results of {\sc dtunuc}
simulations are in perfect agreement with the measurements. This check
is particularly important as it stands within the projectile energy
range where most cosmic antiprotons are produced.

                         %###############%
\subsubsection{Comparison with analytical parameterization from 
Mokhov \& Nikitin}
Taking into account the qualitative predictions of the Regge phenomenology and
partons model, Mokhov \& Nikitin (1977) derived a parametrized
inclusive cross-section for $p + A \rightarrow \bar{p} +X$:
$$
{\left( E \frac{d^3\sigma}{d^3p}\right)}_{inv} = \sigma_{abs} C_1
^{b(p_T)}(1-x')^{C_2}\exp(-C_3 x') \Phi (p_T)\;\;,
$$
$$
\Phi(p_T) = \exp(-C_4p_T^2) + C_5 \frac{\exp(-C_6 x_T)}{{(p_T^2 +
\mu^2)}^4}\;\;,
$$
where
$$
b(p_T)=\left\{\begin{array}{r}
b_op_T  \hbox{ if } p_T\leq \Gamma\;\;;\\
b_o\Gamma \hbox{ otherwise }\;\;.
\end{array}
\right.
$$
S is the invariant mass of system, $p_T$ is the transverse momentum, $x_T
\approx{2p_T}/{\sqrt S}$, $x'=
{E^*}/{E^*_{max}}$, ${E^*}$ and $E^*_{max}$ are the total energy of the
inclusive particle in the center of mass frame and its maximum possible value.
The parameters $C_1$ to $C_6$, $b_0$, $\mu^2$ and $\Gamma$ were not taken as
given in Kalinovskii et al. (1989) but were re--fitted using an 
extensive set of
experimental data leading to a better $\chi^2$ \cite{huang}.

Contrary to experimental measurements that are only available for
a small number of given
energies, this analytical approach allows a useful comparison with
{\sc dtunuc} cross-sections. The resulting spectrum has therefore been
propagated using the model described in Section \ref{section:la}
and the results
are in excellent agreement. The {\sc dtunuc} approach was nevertheless
preferred since the Mokhov--Nikitin formula was fitted on rather heavy
nuclei, and its use for p--He, He--p and He--He collisions would therefore
require a substantial extrapolation.
                         %##################%
\subsubsection{Results for the antiprotons}
The exclusive cross-section for antiproton production
${d\sigma^{i,j}}/{dE_{\bar{p}}}\left(E_{\bar{p}},E_i\right)$,
is obtained by multiplying the total inelastic cross-section of
the considered reaction and the antiproton multiplicity interaction given
by {\sc dtunuc}.

\begin{figure}[p]
$$
\epsfxsize=17cm
\epsfysize=16cm
\epsfbox{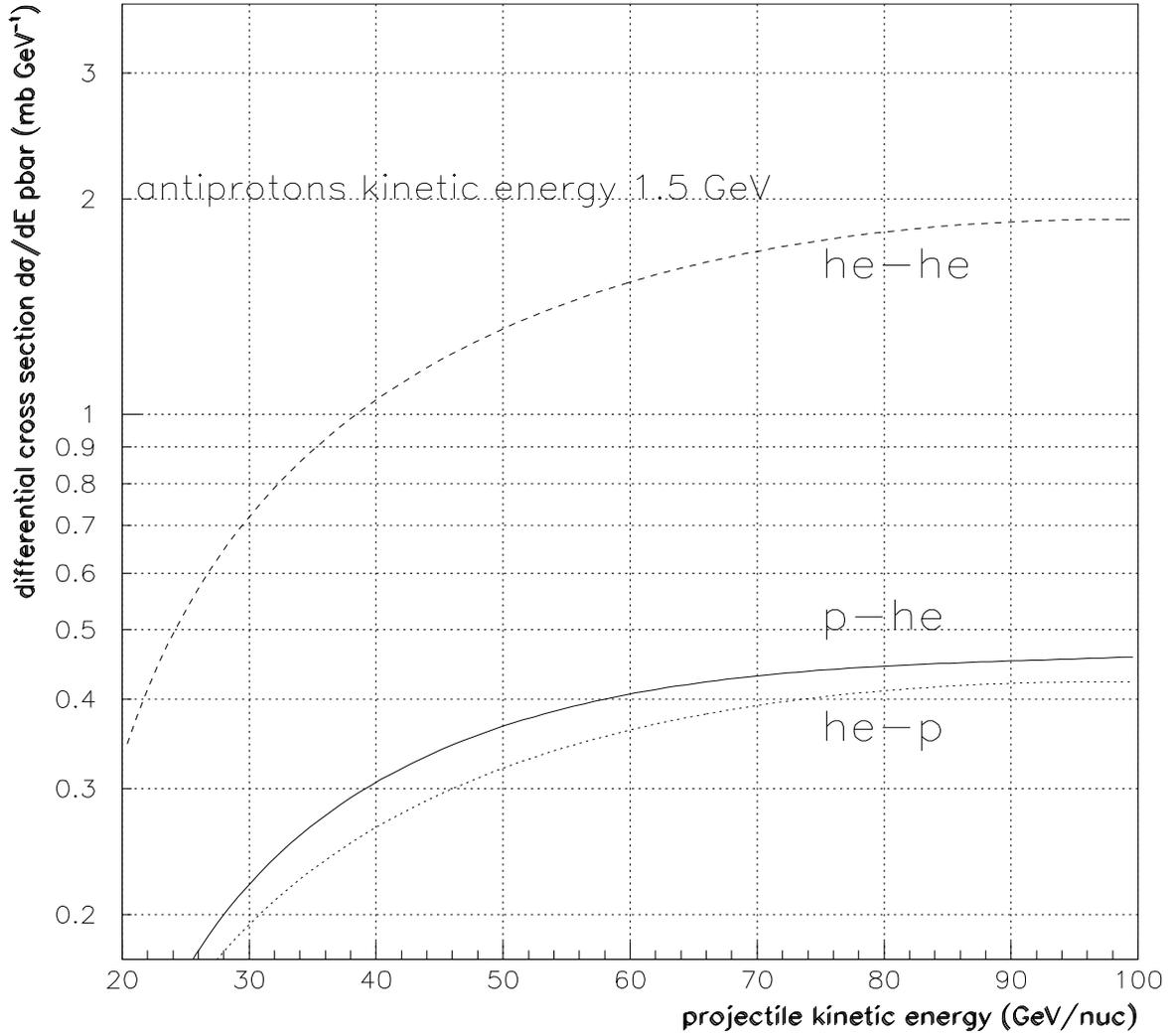}
$$
\caption{From top to bottom: antiproton differential
production cross-section in He--He, p--He and He--p
reactions for antiprotons kinetic energy 1.5 GeV,
as obtained with {\sc dtunuc} simulations.}
\label{fig:pbar}
\end{figure}

This approach is time consuming since the cross-section is quite low and
a large number of events must be generated to reach acceptable statistical
uncertainties. The sampling points were chosen to be distributed on a
logarithmic scale between 7 GeV (threshold) and 10 TeV per nucleon 
for the projectile nucleus
and  extrapolations rely on polynomial fits. The antiproton kinetic energy
was varied from 0.1 GeV to 100 GeV. Figure \ref{fig:pbar} gives some 
examples of
differential antiproton production cross-sections as obtained from 
{\sc dtunuc}.

%%%%%%%%%%%%%%%%%%%%%%%%%%%%%%%%%%%%%%%%%%%%%%%%%%%%%%%%%%%%%%%%%%%%%%%%%%%%
\section{Tertiary contribution}
\label{section:la}

Once they have been created, antiprotons may interact with the
interstellar material in three different ways.
First, they may undergo elastic scatterings on galactic hydrogen.
The cross-section for that reaction has been shown to peak in the
forward direction \cite{eisenhandler}
so that the corresponding antiproton energy loss is negligible.
Antiprotons are not perturbed by these elastic scatterings as
they survive them while their energy does not change.
They may also annihilate on interstellar protons. This process
dominates at low energy, and  its cross-section is given in Tan \& Ng (1983).
Last but not least, antiprotons may survive inelastic scatterings
where the target proton is excited to a resonance. Antiprotons
do not annihilate but lose a significant amount of their kinetic
energy.
Both annihilations and non--annihilating interactions contribute
to the inelastic antiproton cross-section so that
\begin{equation}
\sigma^{\bar{p}p}_{non-ann} \; = \;
\sigma^{\bar{p}p}_{ine} \, - \, \sigma^{\bar{p}p}_{ann} \;\; ,
\end{equation}
where $\sigma^{\bar{p}p}_{ine}$ is parametrized as in Tan \& Ng (1983).

For an antiproton kinetic energy $T_{\bar{\rm p}} \gtrsim 10$ GeV,
the Tan \& Ng parameterization of $\sigma^{\bar{p}p}_{ann}$ --
which is based on experimental data -- is no longer valid.
The annihilation cross-section tends furthermore to be small at high energy.
In any case, the antiproton inelastic but non--annihilating interaction cross
section becomes equal to the total proton inelastic cross-section
\begin{equation}
\sigma^{\bar{p}p}_{non-ann} \equiv
\sigma^{pp}_{ine}\;\;.
\end{equation}
The low and high energy relations for $\sigma^{\bar{p}p}_{non-ann}$
do match for an antiproton kinetic energy of $T_{\bar{\rm p}} = 13.3$ GeV.

The energy distribution of antiprotons that have undergone
an inelastic but non--annihilating interaction has not been
measured. It has been assumed here to be similar to the proton
energy distribution after p--p inelastic scattering. An
impinging antiproton with kinetic energy $T'_{\bar{\rm p}}$
has then a differential probability of
\begin{equation}
{\displaystyle \frac{dN_{\bar{p}}}{dE_{\bar{p}}}} \; = \;
{\displaystyle \frac{1}{T'_{\bar{\rm p}}}}
\end{equation}
to end up with the final energy $E_{\bar{p}}$.
That reaction leads to the flattening of their energy
spectrum as the high--energy species of the peak that sits
around a few GeV may replenish the low--energy part of the energy
distribution. The corresponding source term for these so--called
tertiary antiprotons may be expressed as
\begin{eqnarray}
q_{\bar{p}}^{ter}(r , E_{\bar{p}}) & = &
{\displaystyle \int_{E_{\bar{p}}}^{+ \infty}}
\, {\displaystyle
\frac{d \sigma_{\bar{p} \, H \to \bar{p} \, X}}{dE_{\bar{p}}}}
\left\{ E'_{\bar{p}} \to E_{\bar{p}} \right\} \, n_{H} \,
v'_{\bar{p}} \; N^{\bar{p}}(r , E'_{\bar{p}}) \; dE'_{\bar{p}}
\nonumber \\
& - & \;\;\;\;\; \sigma_{\bar{p} \, H \to \bar{p} \, X}
\left\{ E_{\bar{p}} \right\}
\, n_{H} \, v_{\bar{p}} \; N^{\bar{p}}(r , E_{\bar{p}}) \;\;.
\end{eqnarray}
Since the differential cross-section is given by
\begin{equation}
{\displaystyle
\frac{d \sigma_{\rm \bar{p} \, H \to \bar{p} \, X}}{dE_{\bar{\rm p}}}}
\; = \; {\displaystyle
\frac{\sigma^{\bar{p}p}_{non-ann}}{T'_{\bar{\rm p}}}} \;\;,
\end{equation}
the tertiary production term translates into
\begin{equation}
\label{Belle_et_Sebastien}
q_{\bar{p}}^{ter}(r , E) \; = \; 4 \, \pi \, n_{H} \left\{
{\displaystyle \int_{E}^{+ \infty}} \, {\displaystyle
\frac{\sigma^{\bar{p}p}_{non-ann}(E')}{T'}} \,
\Phi_{\bar{p}}(r , E') \, dE' \; - \;
\sigma^{\bar{p}p}_{non-ann}(E) \, \Phi_{\bar{p}}(r , E) \right\} \;\;.
\end{equation}
The integral over the antiproton energy $E$ of $q_{\bar{p}}^{ter}(E)$
vanishes. This mechanism does not actually create new antiprotons.
It merely redistributes them towards lower energies and tends therefore
to flatten their spectrum. Notice in that respect that the secondary
antiproton spectrum that results from the interaction of cosmic ray
protons impinging on interstellar helium is already fairly flat below
a few GeV. Since it contributes a large fraction to the final result,
the effect under scrutiny here may not be as large as previously thought
\cite{bergstrom99}.

As a matter of fact, antiprotons interact on both the hydrogen and
helium of the Milky--Way ridge. Helium should also be taken into
account in the discussion. As explained in Appendix (\ref{Cornofulgur}),
we have replaced the hydrogen density in relation (\ref{Belle_et_Sebastien})
by the geometrical factor $n_{H}+4^{2/3}\:n_{He}$
for the calculation of the tertiary component.

%%%%%%%%%%%%%%%%%%%%%%%%%%%%%%%%%%%%%%%%%%%%%%%%%%%%%%%%%%%%%%%%%%%%%%%%%%%%
\section{Propagation in a diffusion model}
\label{sec:propagation}

Propagation of cosmic rays can be studied within different
theoretical frameworks, the most popular being the so-called Leaky Box
model and the diffusion model. There is a mathematical equivalence of
these two approaches, which is valid only under special
circumstances. In particular, they lead to different results for low
grammages and for unstable cosmic ray species (see discussion in
Maurin et al. 2001). Our preference for the diffusion model has
several justifications. First, it is a more physical approach, in the
sense that cosmic rays are believed to diffuse in the galactic disk and
halo, which is in disagreement  with the spatial homogeneity
assumed in the Leaky Box. Second, the parameters entering the diffusion models
are related to measurable physical quantities (at least in principle),
like the galactic magnetic field, so that their value could be
cross-checked with independent measurements.
Finally, the diffusion approach is mandatory if one wants to take
primary sources into account, as emphasized in the introduction.

The geometry of the problem used here is a classical cylindrical
box (see for example Webber, Lee, \& Gupta 1992) whose radial extension is
$R=20\unit{kpc}$, with a disk of thickness $2h=200\unit{pc}$
and a halo of half--height $L$ lying in the interval $[1-15]\unit{kpc}$.
Sources and interactions with matter are confined to the thin disk and
diffusion which occurs throughout disc and halo with the same strength
is independent of space coordinates. The solar system is located in
the galactic disc ($z=0$) and at a centrogalactic distance
$R_\odot=8 \unit{kpc}$ (Stanek \& Garnavich 1998; Alves 2000).
We emphasize that this model is exactly the one that has been used for
the propagation of charged nuclei (Paper I) where it has been described
in details.
For the sake of completeness, we rewrite here the basic ingredients,
and the parameters of the diffusion model we used.
                         %%%%%%%%%%%%%%
      \subsection{The five parameters of the model}
Our model takes into account the minimal known physical processes
thought to be present during the propagation.
Firstly, the diffusion coefficient  $K(E)$
\begin{equation}
           K(E) = K_0 \, \beta \times {\cal R}^\delta
           \label{eq:diffusion_coefficient}
\end{equation}
where the normalisation $K_0$ is expressed in kpc$^2\unit{Myr}^{-1}$
and $\delta$ is the spectral index (${\cal R}=p/Z$ stands for the
particle rigidity). Along with the spatial
diffusion, one has the associated diffusion in energy space represented by a
reacceleration term
  \begin{equation}
    \label{All is for the butter}
K_{EE}(E)=\frac{2}{9}{V_a}^2\frac{E^2\beta^4}{K(E)}\;\;.
\end{equation}
Here $K_{EE}$ stands for the energy diffusion coefficient
which we evaluated in the no--recoil hard sphere scattering
centers approximation. In particular $V_a$ is the alfv\'enic
speed of scatterers responsible of the energetic diffusion.
Next, we allow a constant convective wind directed outward in the $z$
direction. This term is represented by the velocity $V_c$.
Motivation of such forms for the various parameters has been given
in Paper I and will not be repeated here. Last, we have to include
effects of energy losses. Formul{\ae} for the latter are those used
for nuclei with the appropriated charge for an antiproton
(see Paper I).

As a consequence, diffusion model is described with five parameters:
the diffusion coefficient normalization $K_0$ and its power index $\delta$,
the convective galactic wind velocity $V_c$, the Alfv\'enic speed $V_a$,
and finally the halo thickness $L$.

                         %%%%%%%%%%%%%%
\subsection{Configuration of the parameter space used for this analysis}

The values of these parameters are needed to compute
the propagated antiproton flux. They may be extracted from a careful
analysis of charged cosmic ray nuclei data.
This has been done in a previous study (Paper I), where all
the sets of parameters consistent with B/C and sub--Fe/Fe data were
determined.
As the propagation history for all cosmic rays should be similar,
this is thought to be a safe procedure.
In this work, we used the same sets and the same numerical code to propagate
antiprotons, to make sure our treatment is fully consistent with
our previous work and that the results are consistent with nuclei data.
This is in variance with previous works using diffusion models, where
the propagation parameters were extracted from a leaky box analysis of
nuclei.
It should be noticed that some of the sets of parameters are probably 
disfavored by physical considerations. For instance, our models have 
Alfven velocities $V_a$ ranging from $25 \unit{km} \unit{s}^{-1}$ to 
$85 \unit{km} \unit{s}^{-1}$.  
The upper end of this range is too high.
Indeed,  the value of the galactic magnetic 
field ($B \approx 1-2 \unit{$\mu$G}$, see for example Han \& Qiao (1994)
or Rand \& Lyne (1994))
and the plasma density ($\langle n_e \rangle = 0.033 \unit{cm}^{-3}$ 
according to Nordgren, Cordes \& Terzian (1992)) give
$10 \unit{km} \unit{s}^{-1} \lesssim V_a \lesssim 30 \unit{km} 
\unit{s}^{-1}$.
Besides, as mentionned in Paper I, the physical meaning of the value of $V_a$ 
may depend on the assumptions made for the scattering process.
A proportionality coefficient larger than $2/9$ in relation 
(\ref{All is for the butter}) would imply smaller values for $V_a$. 
The following point should also be kept in mind: 
we considered that reacceleration only occured in the thin disk, 
{\em i.e.} in a zone of half-height $h_a=h=100  \unit{pc}$. 
If this process is efficient in a larger zone ($h_a > h$),
the overall effect is unchanged provided that the Alfven velocity is 
scaled down to a lower value as $V_a \propto (L/h_a)^{1/2}$  
(Seo \& Ptuskin 1994). 
In our semi-analytical resolution of the diffusion model, the 
case $h_a \neq h$ cannot be straightforwardly taken into 
account, but the previous conclusion would still hold.
Indeed, we can make the reacceleration zone larger by
increasing the disk thickness $h$, while keeping constant the 
quantity $n_H h$ so that all the other effects are unaffected.
For example, a $h_a = 1 \unit{kpc}$ 
reacceleration zone would lead to Alfven velocities about three times 
smaller so that in the sets of parameters used in this study, $V_a$ 
would range between $\sim 10 \unit{km} \unit{s}^{-1}$ and 
$\sim 30 \unit{km} \unit{s}^{-1}$.
Anyway, we adopt a conservative attitude and we do not apply any cut in our initial 
sets of parameters.

To sum up, we have applied all the configurations
giving a good $\chi^2$ (less than 40 for 26 data points and 5
parameters) in the B/C analysis of Paper I
(see this paper for an extensive description of the nuclei
analysis).
We insist on the fact that none of this parameter is further modified
or adjusted, they are not free parameters.

                         %%%%%%%%%%%%%%
      \subsection{Calculation of the secondary component}

Once the set of diffusion--propagation parameters is chosen as explained
above, evaluation of the
corresponding flux is straightforward. Semi--analytical solution for the
antiproton background
is given in Appendix \ref{sec: sol diffusion model}.
Apart from the propagation, the two other necessary inputs are -- as one can
see from equation (\ref{SECONDARY_1}) -- the measured top of atmosphere H
and He flux discussed in Section \ref{section:trop bien},
and the nuclear processes described in Section \ref{section:ici}
and \ref{section:la}.

To compare our results to experimental data, solar
modulation (the effect of the solar wind on the interstellar flux
crossing the heliosphere) must be taken into account.
We chose to use the so-called force-field approximation, which is used in
most antiproton studies (see last section for a discussion).

In all the subsequent results, the top--of--atmosphere antiproton
flux has been obtained from the interstellar one with a modulation parameter
of $\phi = 250 \unit{MV}$ ($\Phi \equiv Z/A \times \phi = 250 \unit{MV}$),
adapted for a period of minimal solar activity.
This choice is motivated by the comparison to {\sc bess} data
taken during the last solar minimum.

%%%%%%%%%%%%%%%%%%%%%%%%%%%%%%%%%%%%%%%%%%%%%%%%%%%%%%%%%%%%%%%%%%%%%%%%%%%%
\section{Results and uncertainties}
\label{sec:results}

\subsection{Results}

\begin{figure}[p]
\begin{center}
\includegraphics*[bb=55 118 530 635,clip,width=1.\textwidth]
{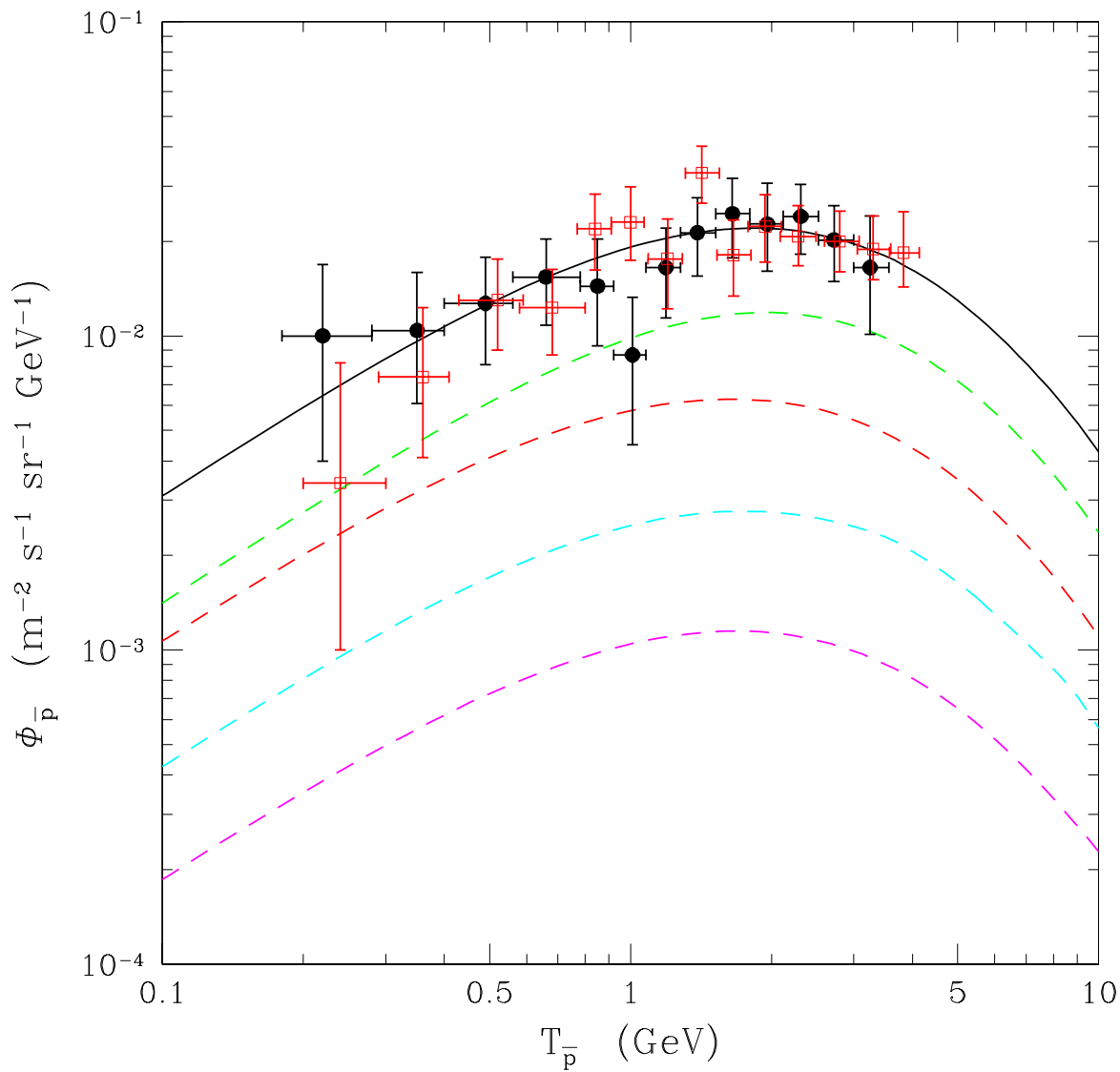}
\caption{Solid line shows the total top-of-atmosphere ({\sc toa})
secondary antiproton spectrum
for the reference set of diffusion parameters (see text for details).
Dashed lines are the contributions to this total flux from various
nuclear reactions (from top to bottom: p--p, p--He, He--p and He--He).
Data points are taken from {\sc bess 95+97} (filled circles) and from
{\sc bess 98}  (empty squares).}
\label{pbar_nuclear_contribution}
\end{center}
\end{figure}

We have calculated the secondary, top--of--atmosphere antiproton
spectrum obtained with the procedure described above.
To begin with, we chose a particular set of diffusion parameters
giving a good fit to the B/C data (see above).
Namely, we have fixed: $K_0/L = 0.0345 \unit{kpc}^2\unit{Myr}^{-1}$,
$L = 9.5 \unit{kpc}$,
$V_c = 10.5 \unit{km}/$s and $V_a = 85.1 \unit{km}/$s.
This set gives the best $\chi^2$ for $\delta$ fixed to $0.6$ and
the resulting antiproton spectrum will be used as a reference in most
subsequent figures.
Fig.~\ref{pbar_nuclear_contribution},
displays this computed antiproton flux along with
experimental data collected by the {\sc bess} spectrometer during two
flights in a
period of minimal solar activity. Circles correspond to the combined 1995 and
1997 data (Orito et al. 2000) and squares to the 1998 ones (Maeno et al. 2000).
The dotted lines represent the contribution to the total flux coming from the
various nuclear reactions: from top to bottom are represented the contribution
of p--p, p--He, He--p and He--He.

First of all, we notice that the calculated spectrum
agrees very well with the {\sc bess} data points.
This strong result gives confidence in our consistent treatment of nuclei and
antiproton propagation.
Second, even if the main production channel is the spallation of cosmic ray
protons over interstellar hydrogen, we see that the contribution of protons
over helium is very important, particularly at low energies (where a
hypothetical primary signature would be expected).
It emphasizes the necessity of having a good
parameterization of the p--He reaction.

In the following sections, we study and quantify all the
uncertainties and possible sources of errors in the
secondary antiproton flux given above.

                    %###################################%
\subsection{Uncertainties from diffusion parameters}
\label{sec:uncertainties}

\begin{figure}[p]
\begin{center}
\includegraphics*[bb=55 118 530 635,clip,width=1.\textwidth]
{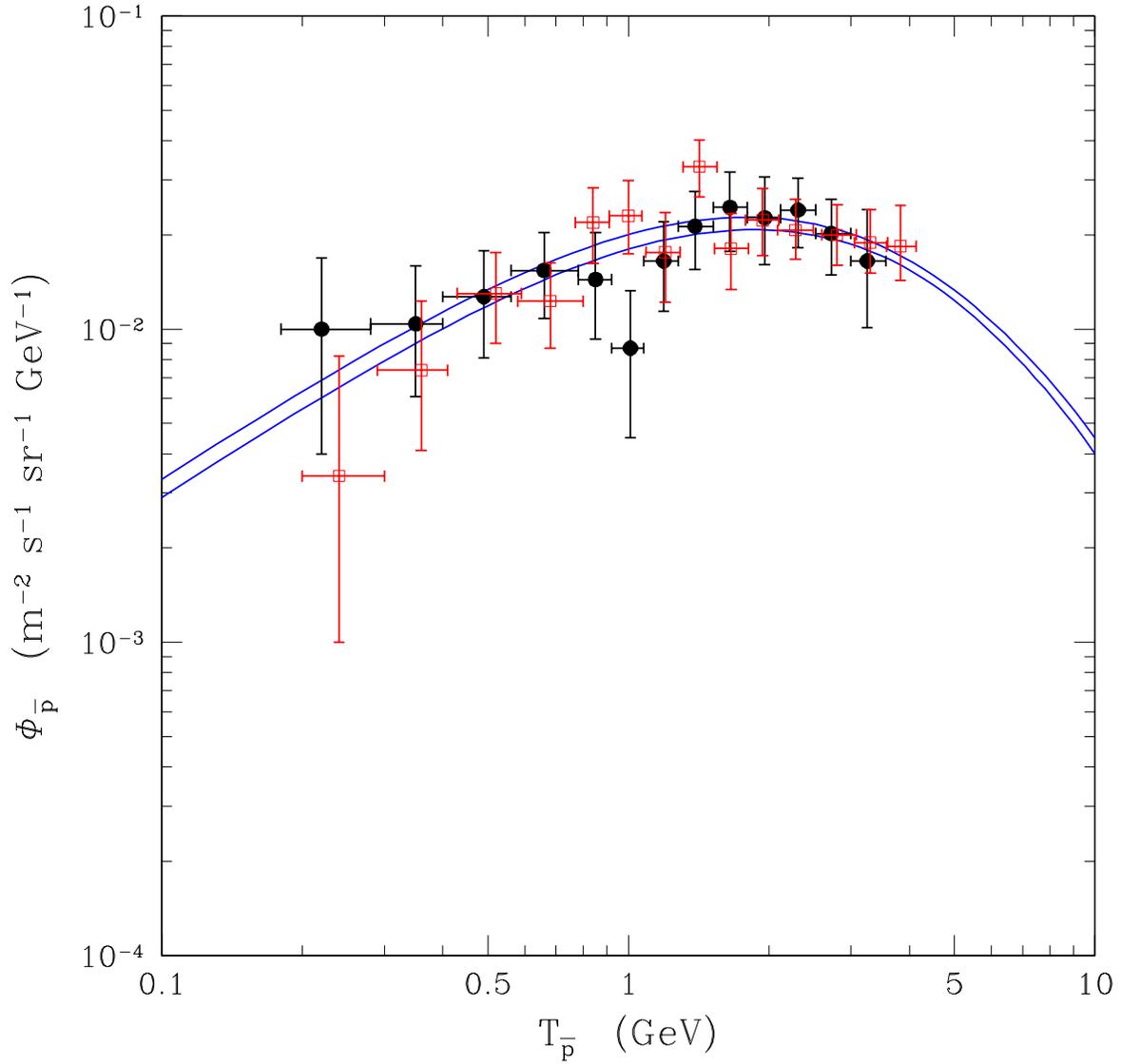}
\caption{This plot shows the envelope of the {\sc toa} antiproton spectra
generated with the sets of diffusion parameters consistent with B/C
and for which $\delta$ has been fixed to 0.6
(data points are the same as in Fig.~\ref{pbar_nuclear_contribution}).}
\label{delta_06}
\end{center}
\end{figure}

\begin{figure}[p]
\begin{center}
\includegraphics*[bb=55 118 530 635,clip,width=1.\textwidth]
{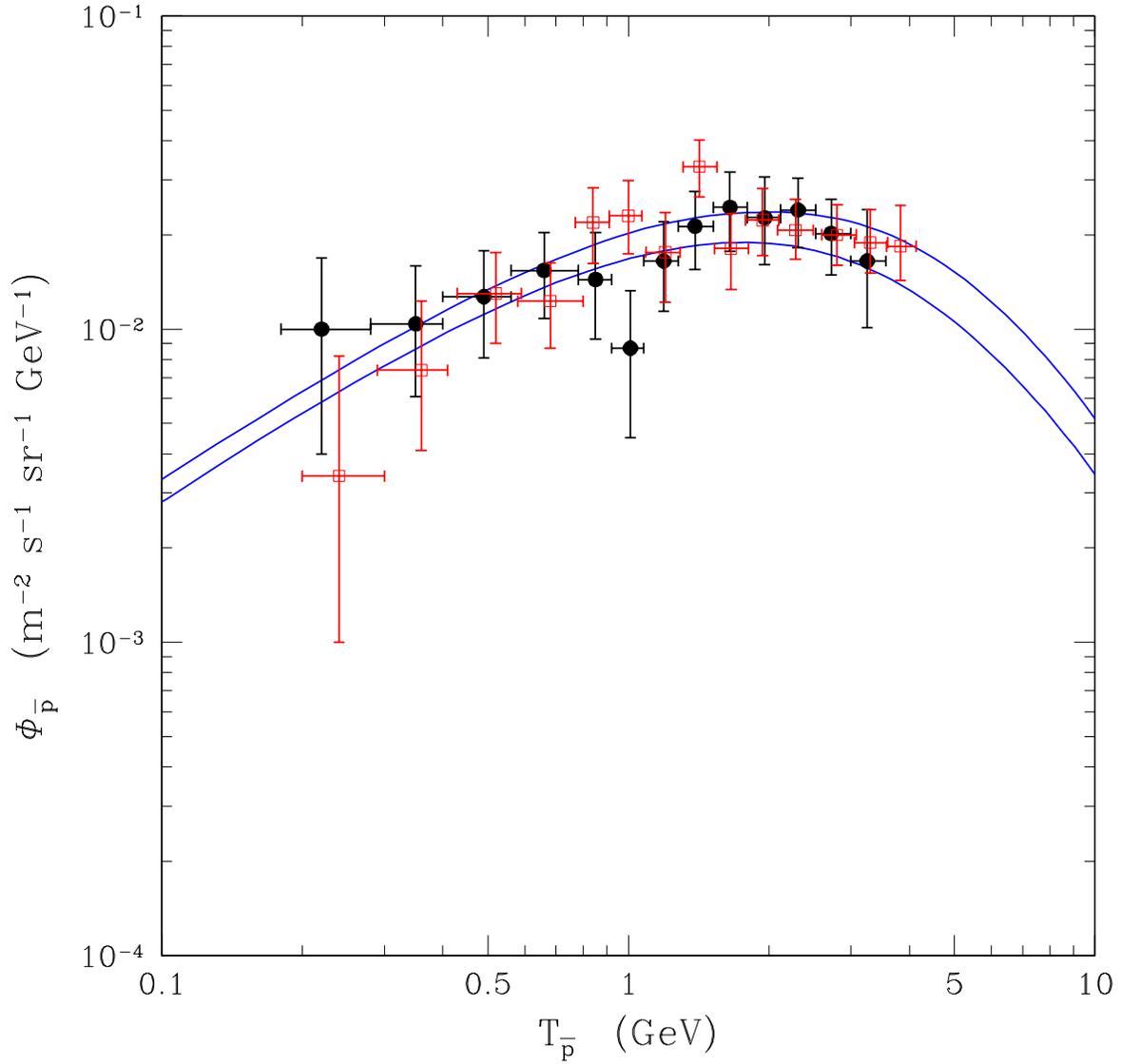}
\caption{Same as previous figure, but where the whole region
of parameter space consistent with B/C has been used (Fig. 7 of Paper I).
The resulting bounds give an estimation of the uncertainty due
to the undeterminacy of the diffusion parameters
(data are the same as in Fig.~\ref{pbar_nuclear_contribution}).}
\label{all_delta}
\end{center}
\end{figure}

The first source of uncertainty comes from the fact
that the propagation parameters are not perfectly known, even if they
are severely constrained by the analysis of B/C experimental results
(Paper I).
A quantitative estimate for this uncertainty is obtained by applying
all the good parameter sets to antiproton propagation.
In a first step, we set the diffusion coefficient spectral index
$\delta$ to 0.6 and allow the four other parameters
($K_0$, $L$, $V_c$ and $V_a$) to vary in the
part of the parameter space giving a good fit to B/C.
The resulting antiproton fluxes are presented in Fig.~\ref{delta_06}.
The two curves represent the minimal and the maximal flux obtain with
this set of parameters.
In a second step, we also let $\delta$ vary in the allowed region of the
parameter space, along with the four other parameters (Fig.7 and 
Fig.8 of Paper I).
As before, the minimal and maximal
fluxes are displayed in Fig.~\ref{all_delta}.
The resulting scatter depends on the energy.
More precisely, it is 9$\%$ from 100 MeV to 1 GeV, reaches a maximum
of 24$\%$ at 10 GeV and decreases to 10$\%$ at 100 GeV.
This gives our estimate of the uncertainties related to diffusion. 
They may be considered as quite
conservative, as the range of allowed parameters could probably be further
reduced by a thorough analysis  of radioactive nuclei
(Donato et al, in preparation) and also by new measurements of stable species.

                    %###################################%
\subsection{Uncertainties from nuclear parameters}

The uncertainties on the antiproton production cross-sections from p-He,
He-p and He-He reactions have been evaluated using the most
extensive set of experimental data available. In addition to those
described in Section \ref{Aurelien deconne}, the average antiproton 
multiplicity in p--p
collisions as measured by Antinucci et al. (1973) has also been checked out.
Finally, measurements from Eichten et al. (1972)
performed by the {\sc cern}-Rome group with the single-arm
magnetic spectrometer \cite{allaby} were taken
into account. They give the Lorentz invariant density (defined as
$2E \delta^2 \sigma / (\sigma_a p^2 \delta p \delta \Omega)$ where $E$ and
$p$ are the laboratory energy and momentum of the produced antiproton and
$\sigma_a$ is the absorption cross-section) as a function of $p$ and of
the production angle $\theta$. A wide range of values from $\theta=17$ mrad
to $\theta=127$ mrad and from $p=4$ GeV to $p=16$ GeV has been explored.

All those measurements have been compared with {\sc dtunuc} results. As
mentionned before, most of them are in excellent agreement with the simulation.
The more important discrepancies were found for high-energy produced
antiprotons in p-Be collisions and for low energy projectile protons in p-p
collisions. This latter point is not surprising as the physical input of
{\sc dtunuc} can hardly be justified for a center of mass energy $\sqrt s <$
10 GeV. In both cases, experimental cross-sections were lower than the
simulated ones. Differences are never larger than a factor of two. To account
for such effects we parameterized {\it maxima} and {\it mimina} cross-sections
as a correction to the computed ones, depending on the projectile and
antiproton energies. The simplest, {\it i.e.} linear, energy variation
was assumed and the slope was chosen to be very conservative with respect
to experimental data.
Finally, it has been checked that changes in the Monte Carlo results
induced by small variations of the input physical parameters remain
within the previously computed errors.

According to Tan \& Ng (1982, 1983),
the uncertainty in the parameterization  of their p--p cross-section
should not exceed 10$\%$. From another point of view, Simon et al. (1998)
have compared two parameterization of the existing data along with the
Monte Carlo model {\sc dtunuc}. They found large discrepancies which
induce a 40\% effect on the antiproton prediction. Nevertheless,
since data are available for that reaction, we think that
the Tan \& Ng parameterization is more reliable than any Monte Carlo.

\begin{figure}[p]
\begin{center}
\includegraphics*[bb=55 118 530 635,clip,width=1.\textwidth]
{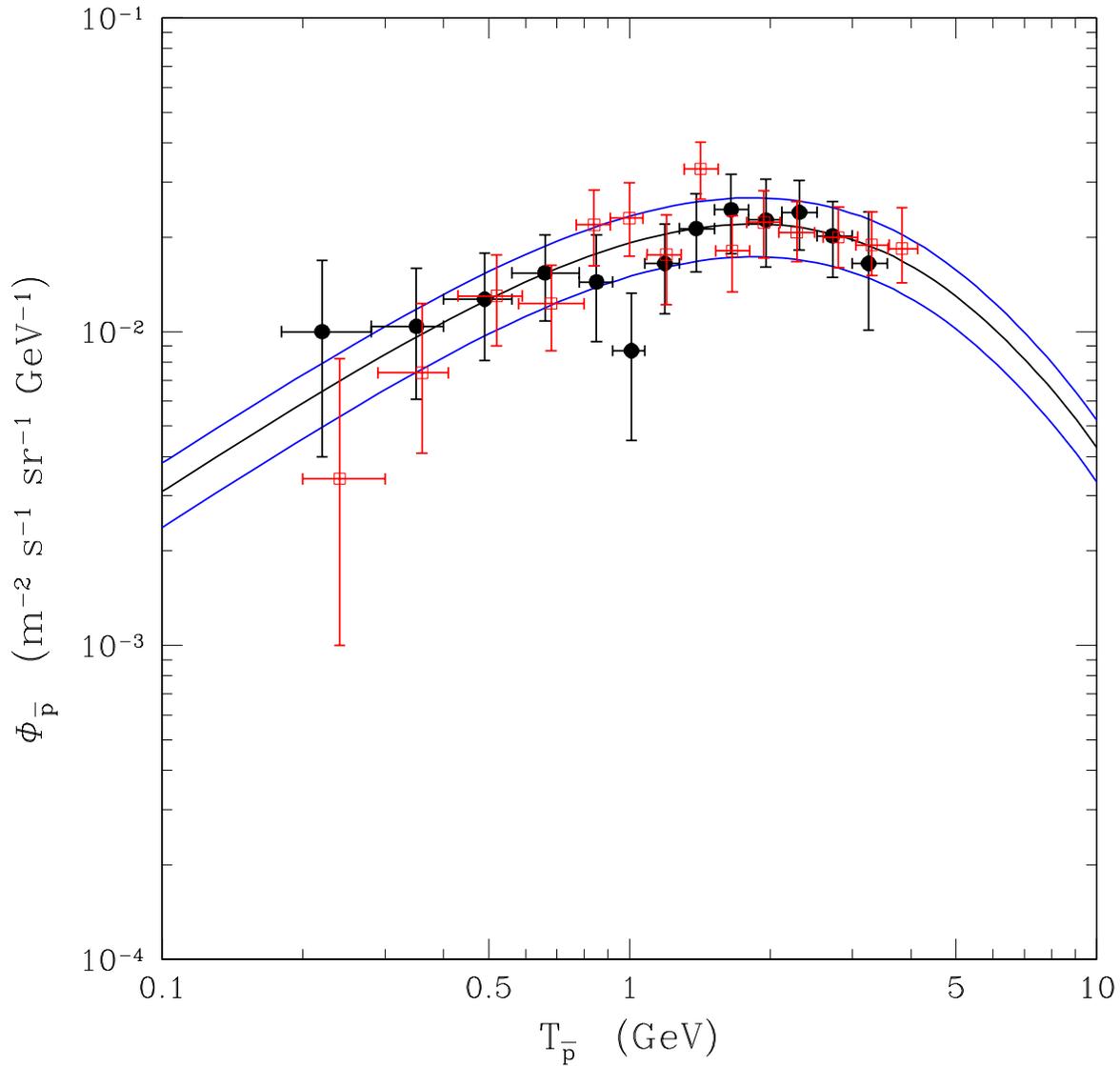}
\caption{In this figure, the {\sc toa} antiproton spectrum has been computed
with extreme values of {\sc dtunuc} nuclear parameters. The central
line is the
reference curve showed in Fig. \ref{fig:pbar}, while upper and lower curves
correspond respectively to the maximum and minimum of the antiproton
production rate.
These two bounds give an estimation of the uncertainty due
to the undeterminacy of the nuclear parameters
(data are the same as in Fig. \ref{pbar_nuclear_contribution}).}
\label{pbar_nuclear_uncertainty}
\end{center}
\end{figure}

In Fig. \ref{pbar_nuclear_uncertainty} we present our estimation of the
uncertainties related to nuclear physics.
The central curve is our reference presented above. The upper one is
obtained with the set of maximal p--He, He--p, He--He cross-sections
while increasing the p--p cross-section by 10$\%$. Similarly, the lower
curve is obtained with the minimal values for these cross-sections
while decreasing the p--p cross-section by 10$\%$. Indeed, such a variation
for p--p has been included for the sake
of completeness even if it modifies the antiproton spectrum only by
a few percents.  As a conclusion, the shift of the upper and the
lower curve with respect to the central one is of the order of
22--25~$\%$ over the energy range 0.1--100 GeV.

Besides these major sources of uncertainties, we have also investigated
the influence of a possible error in the parameterization of the
inelastic non--annihilating cross-section, which gives rise to the
tertiary component. We modified it by 20$\%$, which is thought to be
very conservative. We found that the antiproton spectrum is modified
by less than 1$\%$. In the same line of thought, the effect of
total inelastic plus non-annihilating reactions on interstellar He
is found to be negligible (see discussion in Appendix \ref{Cornofulgur}).

%###################################%
\subsection{Other uncertainties}
There are few other sources of uncertainties.
To begin with, as we discussed in Section \ref{section:trop bien},
primary cosmic ray fluxes (protons and helium)
have been measured with unprecedented accuracy. For the first time, the
induced uncertainties on the antiproton spectrum can be neglected.

Next, the only parameters which have not been varied in the previous
discussion are those related to the description of
the interstellar medium, {\em i.e.}
the densities $n\un{\scriptsize H}$ and $n\un{\scriptsize He}$. In all
the preceding analysis,
these were fixed to $n\un{\sc ism}\equiv n\un{\scriptsize H}+n\un{\scriptsize
He}=1\unit{cm}^{-3}$ and $f\un{\scriptsize He}\equiv 
n\un{\scriptsize He}/n\un{\sc ism}=10\%$
(same as in Paper I). We have tested the sensitivity of our results
to changes in both $n\un{\sc ism}$ and $f\un{\scriptsize He}$.
For this purpose, we
found the new values for the diffusion parameters (for $\delta=0.6$)
giving a good fit to B/C, and applied them to antiprotons.
Varying $f\un{\scriptsize He}$ in the range  $5\% < f\un{\scriptsize
He} <15\%$,
the resulting flux is modified by less than $15\%$ over the whole energy
range. Notice that this range of $f\un{\scriptsize He}$ values can
be considered as very conservative (see discussion in Strong \&
Moskalenko 1998). A more realistic 10\% error on
$f\un{\scriptsize He}$ (i.e. $0.9 \%< f\un{\scriptsize He} <1.1\%$)
would lead
to a few \% error on the antiproton spectrum.
Alternatively, varying $n\un{\sc ism}$ from $0.8$ to
$1.2\unit{cm}^{-3}$, the resulting flux is modified by less
than $0.5\%$ over the whole energy range.
To sum up, the only contributing errors are from the helium fraction
$f\un{\scriptsize He}$ through the dependence of antiproton production on
corresponding cross-sections.

\begin{figure}[p]
\begin{center}
\includegraphics*[bb=55 118 530 635,clip,width=1.\textwidth]
{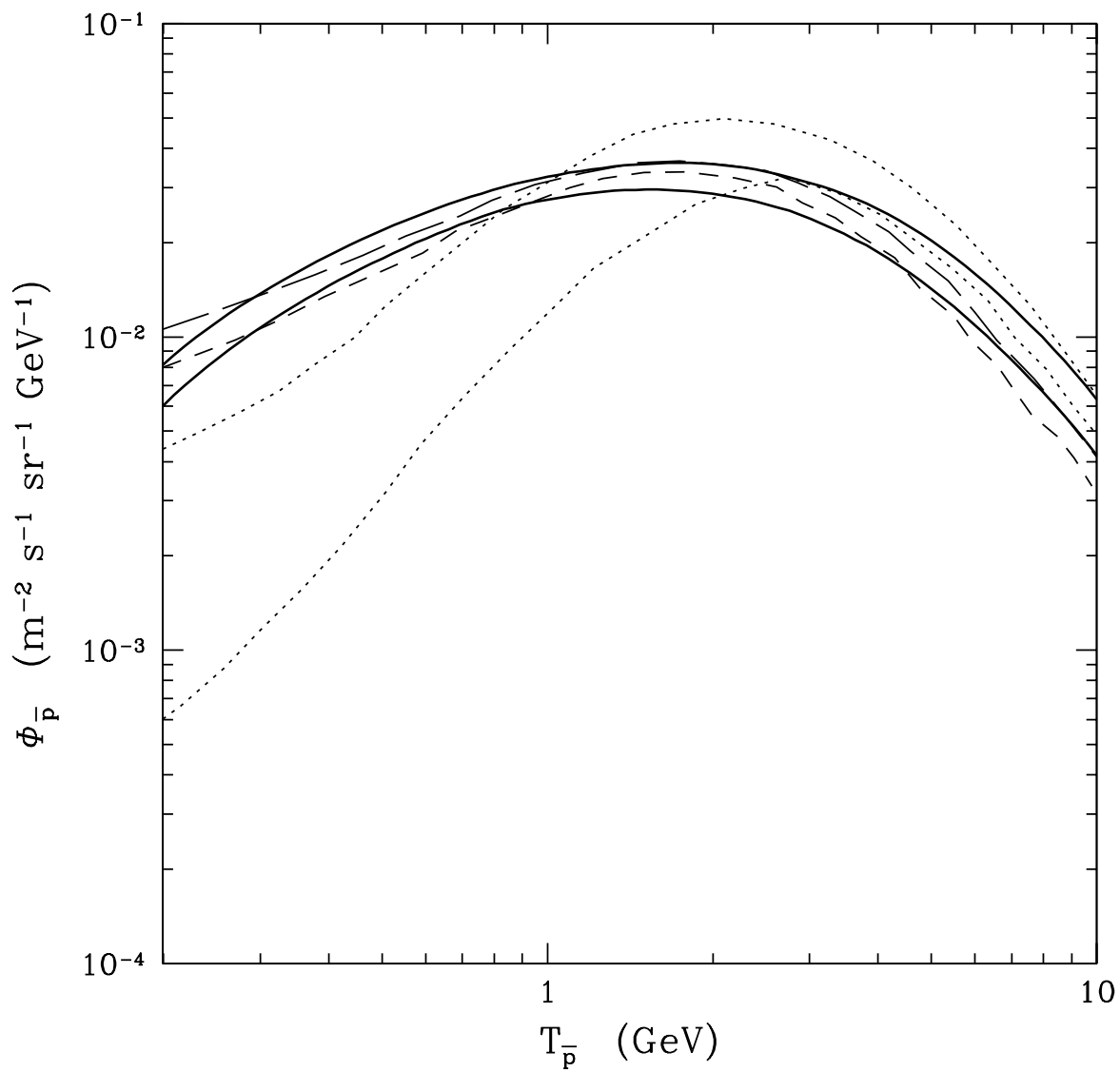}
\caption{Comparison of our interstellar spectra
(thick solid lines indicate the lower 
and upper band due to the uncertainties in the propagation parameters)
with other published antiproton spectra. Dotted lines are lower and 
upper values from Simon et al. (1998), short dashed line is from Bieber et 
al. (1999) and long dashed line is from Moskalenko et al. (2001).}
\label{comparison_other_works}
\end{center}
\end{figure}

Finally, solar modulation induces some uncertainty. This
problem  is still debated, and a rigorous treatment of this effect is
beyond the scope of this paper (see for example Bieber et al.
1999 for a recent analysis). However, in a ``force-field" approximation,
a general feature is that the steeper the spectrum, the greater the effect.
Our antiproton spectra being rather flat, we do not expect them to be
dramatically affected by a change in the modulation parameter.
Anyway, this local effect is decorrelated from the propagation history.
Solar modulation -- which is the last energetic modification
suffered by an incoming galactic cosmic ray -- can thus be treated
completely independently from the above analysis.
Figure~\ref{comparison_other_works} shows our demodulated spectra together 
with other interstellar published spectra
(Simon et al. 1998, Bieber et al. 1999, Moskalenko et al. 2001)

%%%%%%%%%%%%%%%%%%%%%%%%%%%%%%%%%%%%%%%%%%%%%%%%%%%%%%%%%%%%%%%%%%%%%%%%%%%%

\section{Conclusions}
We have computed cosmic antiproton fluxes in the framework of a
two-zone diffusion model taking into account galactic wind, stochastic
reacceleration and energy losses.
The propagation parameters have been chosen according to Maurin et al. (2001),
as to be in agreement with cosmic ray nuclei data.
The annihilating as well as the inelastic non-annihilating (tertiary) p--p
reactions have been taken into account.
The p--p, He--p, p--He and He--He nuclear reaction have also been included
and the relevant cross-sections have been computed using
the Monte Carlo program {\sc dtunuc}.
The latest measured values for cosmic protons and helium fluxes from {\sc ams}
and {\sc bess} have been considered.

The results may be summarized as follows.
First, the values of all the inputs being either extracted from the
analysis of nuclei (diffusion parameters $\delta$, $L$, $K_0$, $V_c$
and $V_a$) or measured
(proton and helium fluxes), all the cosmic antiproton fluxes naturally
coming out of the calculation are completely contained within the experimental
error bars of {\sc bess} data.

The other strong conclusion is that all possible sources of uncertainties 
have been derived. They have been significantly improved with respect 
to the previous gross estimates.

In particular, those related to propagation range between 10\%
and 25\%, depending on which part of the spectrum is considered,
and those related to nuclear physics are below 25 \%.
We emphasize that the uncertainties related to propagation will
probably be further reduced by a more complete study of cosmic ray
nuclei, in particular by focusing on the radioactive species.
We also note that more accurate data on cosmic ray nuclei fluxes
would give better constraints on the diffusion parameters, which in
turn would translate into lower uncertainties on antiprotons fluxes.
The major remaining uncertainties come from nuclear physics and are
already comparable to experimental error bars. As antiproton spectrum
measurements should better in the near future, antiproton studies could
be limited by nuclear undeterminacies. Further work and especially new
measurements of antiproton production in the p--He channel would be of great
interest.

                                 %%%%%%%%%%%%
\section*{Acknowledgments}
We thank the anonymous second referee for useful comments on 
reacceleration.
F.D. gratefully acknowledges a fellowship by the Istituto Nazionale di Fisica
Nucleare. We also would like to thank the French Programme National de
Cosmologie for its financial support.  Finally, we are grateful to S. 
Roesler who
provided us with {\sc dtunuc} and was very helpful in answering our questions.
%%%%%%%%%%%%%%%%%%%%%%%%%%%%%%%%%%%%%%%%%%%%%%%%%%%%%%%%%%%%%%%%%%%%%%%%%%%%%%%%%%%%
%%%%%%%%%%%%%%%%%%%%%%%%%%%%%%%%%%%%%%%%%%%%%%%%%%%%%%%%%%%%%%%%%%%%%%%%%%%%%%%%%%%%
\newpage
\addtocontents{toc}{\protect\newpage}
%\addtocontents{toc}{\hspace{-\parindent}{\Huge\bf Annexes}}
%\addtocontents{toc}{\protect\vspace{4ex}}
\addcontentsline{toc}{part}{Annexes}
\addtocontents{toc}{\protect\vspace{2ex}}

\appendix
%%%%%%%%%%%%%%%%%%%%%%%%%%%%%%%%%%%%%%%%%%%%%%%%%%%%%%%%%%%%%%%%%%%%%%%%%%%%%%%%%%%%
\section{Solution for the secondary antiprotons}
\label{sec: sol diffusion model}

We summarize in this annex the salient features of our derivation
of the spallation antiproton energy spectrum. The propagation of
cosmic--rays throughout the galaxy is described with a two--zone
effective diffusion model which has been thoroughly discussed in a
preceeding analysis (Paper I). The Milky--Way is pictured
as a thin gaseous disk with radius $R = 20$ kpc and thickness
$2 h = 200$ pc where charged nuclei are accelerated and scatter
on the interstellar gas to produce in particular secondary
antiprotons. That thin ridge is sandwiched by two thick confinement
layers. The effective diffusion of cosmic--rays throughout the galactic
magnetic fields occurs uniformly within the disk and halo with the
same strength. Furthermore, we consider here a constant wind $V_c$
in the $z$ direction. The associated adiabatic losses take place
in the disk only.

\subsection{High energy limit}
\label{sec:HEL}

As compared to the cosmic--ray nuclei on which the analysis
of Paper I has focused, antiprotons have the same propagation
history but differ as regards their production. The space--energy
density $N^{\bar{p}}$ is related to the antiproton flux through
\begin{equation}
\Phi_{\bar{p}}(r , E) \; = \;
{\displaystyle \frac{1}{4 \, \pi}} \, v_{\bar{p}}(E) \, N^{\bar{p}}(r , E)
\;\; .
\label{definition_flux}
\end{equation}
As explained in Paper I -- see in particular their
equation~(A1) -- the density $N^{\bar{p}}$ satisfies the
relation
\begin{eqnarray}
2 \, h \, \delta(z) \, q_{\bar{p}}^{sec}(r,0,E) & = &
2 \, h \, \delta(z) \, \Gamma^{ine}_{\bar{p}} \, N^{\bar{p}}(r,0,E)
\; + \; \nonumber \\
& + &
\left\{
V_{c} \frac{\partial}{\partial z} \, - \,
K \left(
{\displaystyle \frac{\partial^{2}}{\partial z^{2}}} +
\frac{1}{r} {\displaystyle \frac{\partial}{\partial r}}
\left( r {\displaystyle \frac{\partial}{\partial r}} \right)
\right) \right\} \, N^{\bar{p}}(r,z,E) \;\; ,
\label{EQUATION_GENERALE}
\end{eqnarray}
as long as steady state holds. Diffusion and convection have
been included. Inelastic interactions on interstellar atoms are
described through the collision rate $\Gamma^{ine}_{\bar{p}}$ which
will be discussed in more detail together with tertiary antiprotons.
The antiproton density
\begin{equation}
N^{\bar{p}}(r,z,E) \; = \;
{\displaystyle \sum_{i=1}^{\infty}} \, N^{\bar{p}}_{i}(z,E) \,
J_{0} \left( \zeta_{i} {\displaystyle \frac{r}{R}} \right) \;\; ,
\label{BE_antiproton_density}
\end{equation}
and the secondary source term
\begin{equation}
q_{\bar{p}}^{sec}(r,0,E) \; = \;
{\displaystyle \sum_{i=1}^{\infty}} \, q_{\bar{p} \; i}^{sec}(E) \,
J_{0} \left( \zeta_{i} {\displaystyle \frac{r}{R}} \right) \;\; ,
\label{BE_antiproton_source}
\end{equation}
may be expanded over the orthogonal set of Bessel functions
$J_{0}(\zeta_{i} x)$ where $\zeta_{i}$ stands for the ith zero of
$J_{0}$ while $i = 1 \dots \infty$. The boundary condition
$N^{\bar{p}} \equiv 0$ is therefore readily ensured for $r = R$.
The Bessel transform of the antiproton density has vertical dependence
\begin{equation}
N^{\bar{p}}_{i}(z,E) \; = \; N^{\bar{p}}_{i}(E) \;
\exp \left\{ {\displaystyle \frac{V_{c} \, z}{2 \, K}} \right\} \;
\left\{
{\sinh \left\{ {\displaystyle \frac{S_{i}}{2}} \, \left( L - z \right)
\right\}} \, / \,
{\sinh \left\{ {\displaystyle \frac{S_{i}}{2}} \, L \right\}} \right\}
\;\; ,
\end{equation}
where the quantity $S_{i}$ is defined as
\begin{equation}
S_{i} \equiv \left\{
{\displaystyle \frac{V_{c}^{2}}{K^{2}}} \, + \,
4 {\displaystyle \frac{\zeta_{i}^{2}}{R^{2}}}
\right\}^{1/2} \;\; .
\end{equation}
Solving equation~(\ref{EQUATION_GENERALE}) with the help of the
Bessel expansions~(\ref{BE_antiproton_density}) and
(\ref{BE_antiproton_source}) leads to the simple relation
\begin{equation}
N^{\bar{p}}_{i}(E) \; = \;
{\displaystyle \frac{2 \, h}{A^{\bar{p}}_{i}}} \;
q_{\bar{p} \; i}^{sec}(E) \;\; ,
\label{solution_HEL}
\end{equation}
that mostly holds at high energy -- say above $\sim 100$ GeV --
where energy losses and diffusive reacceleration do not play
any major role. The coefficients $A^{\bar{p}}_{i}$ are given by
\begin{equation}
A^{\bar{p}}_{i}(E) \equiv 2 \, h \, \Gamma^{ine}_{\bar{p}}
\; + \; V_{c} \; + \; K \, S_{i} \,
{\rm coth} \left\{ {\displaystyle \frac{S_{i} L}{2}} \right\}
\;\; .
\label{definition_Ai}
\end{equation}
Notice that the diffusion coefficient $K$ -- that comes into
play in the definition of $S_{i}$ and therefore of $A^{\bar{p}}_{i}$
-- essentially depends on the rigidity. One should keep in mind that
the relationship between $K$ and the energy per nucleon may actually
depend on the nuclear species at stake through the average charge per
nucleon $Z/A$.

\noindent
Secondary antiprotons are produced by the spallation reactions
of high--energy cosmic--ray protons and helium on the interstellar
material of the Milky Way ridge at $z = 0$. The source term
\begin{equation}
q_{\bar{p}}^{sec}(r,E) \; = \;
{\displaystyle \int\un{Threshold}^{\infty}} \,
{\displaystyle \frac{d \sigma}{dE}}
\left\{ a(E') + b \rightarrow \bar{p}(E) \right\} \, n_{b} \,
\left\{ 4 \, \pi \, \Phi_{a}(r,E') \right\} \, dE'
\label{SECONDARY_1}
\end{equation}
corresponds to particles $a$ -- protons or helium -- impinging
on atoms $b$ -- hydrogen or helium -- at rest. Four different
production channels need therefore to be considered depending on
the nature of the cosmic--rays and of the stellar gas. Proton--proton
collisions are discussed in Section~(\ref{pp_interaction}) whereas
interactions that involve at least a helium nucleus are reviewed in
Section~(\ref{ab_interaction}). Bessel expanding relation~(\ref{SECONDARY_1})
leads to
\begin{equation}
q_{\bar{p} \; i}^{sec}(E) \; = \;
{\displaystyle \int\un{Threshold}^{\infty}} \,
{\displaystyle \frac{d \sigma}{dE}}
\left\{ a(E') + b \rightarrow \bar{p}(E) \right\} \, n_{b} \,
v' \, N^{a}_{i}(E') \, dE' \;\; .
\label{SECONDARY_2}
\end{equation}

\noindent
The primary species $a$ are accelerated in the galactic disk so that
their own production rate may be expressed as
\begin{equation}
q_{a} \left( r , z , E \right) \; = \;
{2 \, h \, \delta(z)} \, q_{a} \left( r , 0 , E \right)
\propto q_{a}^{tot}(E) \, f(r) \;\; ,
\end{equation}
where $q_{a}^{tot}(E)$ denotes the global galactic production rate of
particles -- protons or helium -- with energy $E$ in the energy bin $dE$.
We have assumed here that the energy dependence of that production rate
could be disentangled from its distribution $f(r)$ along the galactic
disk.
The bulk of the secondary antiproton production takes place for a
typical energy of the impinging species of $E \sim 20 - 30$ GeV/n.
Note also that the primary fluxes $\Phi_{a}(E)$ are monotonically
decreasing with the energy $E$. Both energy losses and diffusive
reacceleration have therefore a negligible effect on the spectra
$\Phi_{a}$. We readily infer that the Bessel transform $N^{a}_{i}$
may be expressed as
\begin{equation}
N^{a}_{i}(E) \; = \;
{\displaystyle \frac{q_{i}}{A^{a}_{i}}} \; q_{a}^{tot}(E) \;\; ,
\end{equation}
where the coefficients $A^{a}_{i}$ are given by a relation similar to
(\ref{definition_Ai}) whereas the quantities $q_{i}$ are defined as
\begin{equation}
q_{i} \; = \; {\displaystyle \frac{1}{\pi R^{2}}} \,
{\displaystyle \frac{1}{J_{1}^{2} \left( \zeta_{i} \right)}} \,
\left\{ {\displaystyle \int_{0}^{1}} \, u \, du \,
J_{0} \left\{ \zeta_{i} u \right\} \, f \left\{ r = u R \right\} \right\} \,
\left\{ {\displaystyle \int_{0}^{1}} \, u \, du \, f \left\{ r = u R \right\}
\right\}^{-1} \;\; .
\label{q_i}
\end{equation}
The cosmic--ray flux $\Phi_{a}$ may be determined everywhere as it is
related to the Bessel transform $N^{a}_{i}$ through relations similar
to~(\ref{definition_flux}) and (\ref{BE_antiproton_density}). The
cosmic--ray flux $\Phi_{a}$ scales in particular with the global
galactic production rate $q_{a}^{tot}$. This allows to determine
the latter by imposing that the interstellar proton and helium
fluxes at the solar system do actually match the observations.

\subsection{Full solution without tertiaries}

Forgetting for a while that the inelastic collisions of antiprotons
with the interstellar gas may be disentangled into annihilating and
non--annihilating interactions, we have to modify
relation~(\ref{EQUATION_GENERALE}) so as to take into account now the
energy losses as well as diffusive reacceleration. This is straightforward
since those processes take place only in the disk and not in the halo.
Once again, following the procedure described in Paper I, one gets
the differential equation
\begin{equation}
A^{\bar{p}}_{i} \, N^{\bar{p}}_{i} \; + \; 2 \, h \, \partial_{E}
\left\{ b^{\; \bar{p}}_{loss}(E) \, N^{\bar{p}}_{i} \, - \,
K^{\; \bar{p}}_{EE}(E) \, \partial_{E} N^{\bar{p}}_{i} \right\}
\; = \; 2 \, h \, q_{\bar{p} \; i}^{sec}(E) \;\; ,
\label{SOLUTION NON RELATIVISTE}
\end{equation}
where $b^{\; \bar{p}}_{loss}$ and $K^{\; \bar{p}}_{EE}$ stand
respectively for the energy losses and the diffusion in energy.

\subsection{Full solution with tertiaries}
\label{Cornofulgur}

We have seen that the source term for tertiaries is
\begin{equation}
q_{\bar{p}}^{ter}(r , E) \; = \; 4 \, \pi \, n_{H} \left\{
{\displaystyle \int_{E}^{+ \infty}} \, {\displaystyle
\frac{\sigma^{\bar{p}p}_{non-ann}(E')}{T'}} \,
\Phi_{\bar{p}}(r , E') \, dE' \; - \;
\sigma^{\bar{p}p}_{non-ann}(E) \, \Phi_{\bar{p}}(r , E) \right\} \;\;.
\label{TERTIARY_1}
\end{equation}
Remembering that the antiproton flux $\Phi_{\bar{p}}$ is related
to the space--energy density $N^{\bar{p}}$ through
equation~(\ref{definition_flux}) and Bessel expanding
relation~(\ref{TERTIARY_1}) leads to
\begin{equation}
q_{\bar{p} \; i}^{ter}(E) \; = \;
{\displaystyle \int_{E}^{+ \infty}} \,
{\displaystyle \frac{\sigma^{\bar{p}p}_{non-ann}(E')}{T'}} \,
n_{H} \, v' \, N^{\bar{p}}_{i}(E') \, dE' \; - \;
\sigma^{\bar{p}p}_{non-ann}(E) \, n_{H} \, v \, N^{\bar{p}}_{i}(E)
\;\; .
\label{TERTIARY_2}
\end{equation}
In the thin disk approximation, that expression needs to be
multiplied by $2 \, h \, \delta(z)$. The Bessel transforms
$N^{\bar{p}}_{i}(z = 0 , E)$ of the antiproton density obey
now the integro--differential equation
\begin{equation}
A^{\bar{p}}_{i} \, N^{\bar{p}}_{i} \; + \; 2 \, h \, \partial_{E}
\left\{ b^{\; \bar{p}}_{loss}(E) \, N^{\bar{p}}_{i} \, - \,
K^{\; \bar{p}}_{EE}(E) \, \partial_{E} N^{\bar{p}}_{i} \right\}
\; = \; 2 \, h \, \left\{
q_{\bar{p} \; i}^{sec}(E) + q_{\bar{p} \; i}^{ter}(E) \right\}
\;\; .
\label{diffusion_E_a}
\end{equation}
Notice that in the definition of the coefficients $A^{\bar{p}}_{i}$,
the rate $\Gamma^{ine}_{\bar{p}}$ should now be replaced by
\begin{equation}
\label{De cadix}
\Gamma^{ann}_{\bar{p}}(E) \; = \; \sigma^{\bar{p}p}_{ann}(E) \,
v_{\bar{p}}(E) \, n_{H} \;\;,
\end{equation}
where annihilations alone are considered. The inelastic non--annihilating
reactions are directly dealt with in the tertiary production term
$q_{\bar{p} \; i}^{ter}$.

Helium should also be taken into
account in our discussion of the annihilations as well as of the
inelastic but non--annihilating interactions which antiprotons undergo
with interstellar matter. As there are no measurements, we have adopted
as an educated guess the geometrical approximation which consists
in scaling the appropriate cross-sections by a factor of $4^{2/3}$
when we deal with helium. In the formul{\ae} (\ref{TERTIARY_2}),
(\ref{TERTIARY_2}) and (\ref{De cadix}), we have therefore replaced 
the hydrogen
density $n_H$ by $(n_H+4^{2/3}\;n_He)$. Such a replacement has little effect.
That overall change in the propagated antiproton spectrum is at most 1\%.

\section{Numerical resolution}

We need now to solve the energy--diffusion equation~(\ref{diffusion_E_a})
for each Bessel order $i$. In the absence of diffusive reacceleration
and energy losses, its solution $N^{\bar{p} \; 0}_{i}$ satisfies
the relation
\begin{equation}
A^{\bar{p}}_{i} \, N^{\bar{p} \; 0}_{i} \; = \; 2 \, h \,
\left\{
q_{\bar{p} \; i}^{sec}(E) + q_{\bar{p} \; i}^{ter}(E)
\right\} \;\;.
\end{equation}
Defining the functions
\begin{equation}
{\cal C}(E) \; = \; {\displaystyle \frac{2 \, h}
{A^{\bar{p}}_{i} \; T}} \;\;,
\end{equation}
and
\begin{equation}
a(E) \; = \; {\displaystyle \frac{K^{\; \bar{p}}_{EE}(E)}{T}} \;\;,
\end{equation}
where $T = E_{\bar{p}} - m_{\bar{p}}$ is the antiproton kinetic
energy, allows us to simplify equation~(\ref{diffusion_E_a}) into
\begin{equation}
u \; + \; {\cal C} \, {\displaystyle \frac{d}{d x}}
\left\{ b^{\; \bar{p}}_{loss} \, u \; - \; a \,
{\displaystyle \frac{d u}{d x}} \right\} \; = \;
u^{0} \;\; ,
\label{diffusion_E_b}
\end{equation}
where $u^{0}$ and $u$ respectively stand for
$N^{\bar{p} \; 0}_{i}$ and $N^{\bar{p}}_{i}$. We can express
relation~(\ref{diffusion_E_b}) on a one--dimensional grid extending
from $x_{\rm inf}$ to $x_{\rm sup}$ with
$x = \ln \left( T / T_{\rm inf} \right)$. We are interested in
kinetic energies extending from $T_{\rm inf} = 100$ MeV up to
$T_{\rm sup} = 100$ GeV. The spacing between two points in energy is
\begin{equation}
\Delta x \; = \; {\displaystyle \frac{1}{N}} \, \ln \left\{
{\displaystyle \frac{T_{\rm max}}{T_{\rm min}}} \right\} \;\;,
\end{equation}
where $N$ has been fixed to 150 in our code.
%#define DIM_TAB_PBAR 150
%
Our resolution method lies on the direct inversion of the algebraic
linear equations that translate relation~(\ref{diffusion_E_b}) on the set
of the $N + 1$ different values of the variable $x$. If $j$ denotes
the point at position
\begin{equation}
x_{\displaystyle j} \; = \; {\displaystyle \frac{j}{N}} \, \ln \left\{
{\displaystyle \frac{T_{\rm max}}{T_{\rm min}}} \right\} \;\; ,
\end{equation}
we get
\begin{equation}
u^{0}_{j} \; = \; A_{j , j-1} \, u_{j-1} \; + \; A_{j , j} \, u_{j} \; + \;
A_{j , j+1} \, u_{j+1} \;\;.
\label{diffusion_E_c}
\end{equation}
The matrix $A$ that connects $u$ to $u^{0}$ has been written here so
as to be tridiagonal. This allows for a fast inversion of the algebraic
equation~(\ref{diffusion_E_c}).

\noindent {\bf $\bullet$} For $0 < j < N$, the tridiagonal matrix $A$
may be written as
\begin{equation}
a_{j} \; = \; A_{j , j-1} \; = \; - \,
{\displaystyle \frac{{\cal C}_{j}}{2 \, \Delta x}} \, b^{\rm ion}_{j-1}
\; - \;
{\displaystyle \frac{{\cal C}_{j}}{{\Delta x}^{2}}} \, a_{j-1/2} \;\; ,
\end{equation}
while
\begin{equation}
b_{j} \; = \; A_{j , j} \; = \; 1 \; + \;
{\displaystyle \frac{{\cal C}_{j}}{{\Delta x}^{2}}} \,
\left( a_{j-1/2} + a_{j+1/2} \right) \;\; ,
\end{equation}
and
\begin{equation}
c_{j} \; = \; A_{j , j+1} \; = \;
{\displaystyle \frac{{\cal C}_{j}}{2 \, \Delta x}} \, b^{\rm ion}_{j+1}
\; - \;
{\displaystyle \frac{{\cal C}_{j}}{{\Delta x}^{2}}} \, a_{j+1/2} \;\; ,
\end{equation}

\noindent {\bf $\bullet$} The boundary $j = 0$ corresponds to the
low energy tip $T_{\rm min} = 100$ MeV where we have implemented
the condition $\ddot{u}(x_{\rm min}) = 0$. This translates into
$\dot{u}_{- 1/2} = \dot{u}_{1/2}$ and leads to the matrix elements
\begin{equation}
a_{0} \; = \; A_{0 , -1} \; = \; 0 \;\; ,
\end{equation}
and
\begin{equation}
b_{0} \; = \; A_{0 , 0} \; = \; 1 \; - \;
{\displaystyle \frac{{\cal C}_{0}}{\Delta x}} \, b^{\rm ion}_{0}
\; + \;
{\displaystyle \frac{{\cal C}_{0}}{{\Delta x}^{2}}} \,
\left( a_{1/2} - a_{-1/2} \right) \;\; ,
\end{equation}
and also
\begin{equation}
c_{0} \; = \; A_{0 , 1} \; = \;
{\displaystyle \frac{{\cal C}_{0}}{\Delta x}} \, b^{\rm ion}_{1}
\; - \;
{\displaystyle \frac{{\cal C}_{0}}{{\Delta x}^{2}}} \,
\left( a_{1/2} - a_{-1/2} \right) \;\; .
\end{equation}

\noindent {\bf $\bullet$} We have finally assumed that both $u$
and $u^{0}$ were equal at the high--energy boundary $j = N$. In this
regime, the energy losses and the diffusive reacceleration should not
affect too much the cosmic--ray energy spectrum. This translates into
the simple conditions
\begin{equation}
a_{N} \; = \; A_{N , N-1} \; = \; 0 \;\; ,
\end{equation}
and
\begin{equation}
b_{N} \; = \; A_{N , N} \; = \; 1 \;\; ,
\end{equation}
whereas, by definition
\begin{equation}
c_{N} \; = \; A_{N , N+1} \; = \; 0 \;\; .
\end{equation}

Inverting a tridiagonal matrix such as $A$ may be potentially
dangerous as Jordan pivoting is not implemented in the standard
resolution scheme. As a matter of fact, energy losses and
diffusive reacceleration lead to a moderate change in the
antiproton spectrum. This translates into the fact that
the matrix $A$ is close to unity. We have nevertheless checked
that our results remained unchanged when Gauss--Jordan inversion
was used (Press et al. 1992).
We have also modified relation~(\ref{diffusion_E_b}) into
the time--dependent equation
\begin{equation}
{\displaystyle \frac{\partial u}{\partial t}} \; + \;
u \; + \; {\cal C} \, {\displaystyle \frac{d}{d x}}
\left\{ b^{\; \bar{p}}_{loss} \, u \; - \; a \,
{\displaystyle \frac{d u}{d x}} \right\} \; = \; 0 \;\;.
\label{diffusion_E_d}
\end{equation}
It may be shown that the static solution $u$ to
equation~(\ref{diffusion_E_b}) also obtains from the superposition
\begin{equation}
u \; = \; {\displaystyle \int_{0}^{+ \infty}} \,
u^{burst}(t) \; dt \;\;,
\label{u_CN_scheme}
\end{equation}
of the reaction $u^{burst}(t)$ to an initial burst
\begin{equation}
u^{burst}(0) \; = \; u^{0}
\end{equation}
taking place at $t = 0$ and subsequently evolving according
to relation~(\ref{diffusion_E_d}). The later equation has also
been solved on a discrete set of $N + 1$ values of the antiproton
kinetic energy while a Crank--Nicholson scheme was implemented.
Once again, the result~(\ref{u_CN_scheme}) is the same as what
the direct inversion of the algebraic set of
relations~(\ref{diffusion_E_c}) gives. We are therefore confident
that our resolution procedure is robust.

The tertiary source term depends on the global antiproton
energy spectrum that is itself determined by the differential
equation~(\ref{diffusion_E_a}). Starting from a trial antiproton
spectrum -- say for instance $N^{\bar{p} \; 0}_{i}$ with only the
secondary production mechanism $q_{\bar{p} \; i}^{sec}(E)$
cranked up -- we invert equation~(\ref{diffusion_E_c}). The new
energy spectrum is used to compute the tertiary source term
$q_{\bar{p} \; i}^{ter}(E)$ through the integral~(\ref{TERTIARY_2}).
We may therefore proceed once again through the same steps and
invert the diffusive reacceleration equation~(\ref{diffusion_E_a}) until
the antiproton spectrum becomes stable. We have actually checked that
convergence obtains after $\sim$ 5 recursions.

%%%%%%%%%%%%%%%%%%%%%%%%%%%%%%%%%%%%%%%%%%%%%%%%%%%%%%%%%%%%%%%%%%%%%%%%%%%%%%%%%%%
%%%%%%%%%%%%%%%%%%%%%%%%%%%%%%%%%%%%%%%%%%%%%%%%%%%%%%%%%%%%%%%%%%%%%%%%%%%%%%%%%%%

%%%%%%%%%%%%%%%%%%%%%%%%%%%%%%%%%%%%%%%%%%%%%%%%%%%%%%%%%%%%%%%%%%%%%%%%%%
%%%%%%%%%%%%%%%%%%%%%%%%%%%%%%%%%%%%%%%%%%%%%%%%%%%%%%%%%%%%%%%%%%%%%%%%%%%

\clearpage

\clearpage

\clearpage

\clearpage

\clearpage

\clearpage

\clearpage

\clearpage

\end{document}